\documentclass{article} 
\usepackage{iclr2025_conference,times}

\usepackage{amsmath,amsfonts,bm}

\def\eqref#1{equation~\ref{#1}}

\def\1{\bm{1}}

\DeclareMathAlphabet{\mathsfit}{\encodingdefault}{\sfdefault}{m}{sl}
\SetMathAlphabet{\mathsfit}{bold}{\encodingdefault}{\sfdefault}{bx}{n}

\usepackage{hyperref}
\usepackage{url}
\usepackage{standalone}
\usepackage{tikz}
\usepackage{multirow}
\usepackage{booktabs}
\usepackage[eulergreek]{sansmath}
\usepackage{pgfplots}
\pgfplotsset{compat=newest}
\usetikzlibrary{patterns}
\usepgfplotslibrary{colorbrewer}
\usepgfplotslibrary{groupplots}
\usepackage{setspace}
\usepackage{xspace}
\usepackage{subfig}
\usepackage{natbib}
\newcommand{\ff}{\textsc{ff}\xspace}
\newcommand{\cv}{\textsc{cv}\xspace}

\title{Cross-cultural value alignment frameworks for responsible AI governance: Evidence from China-West comparative analysis}

\author{Haijiang Liu\thanks{Work done during visiting Ph.D. in the HKUST (GZ).}, Jinguang Gu \\
Wuhan University of Science and Technology\\
Wuhan, China\\
\texttt{\{alecliu, simon\}@ontoweb.wust.edu.cn} \\
\AND
Xun Wu \\
The Hong Kong University of Science and Technology (Guangzhou) \\
Guangzhou, China \\
\texttt{wuxun@hkust-gz.edu.cn} \\
\AND
Daniel Hershcovich\\
University of Copenhagen\\
Copenhagen, Denmark \\
\texttt{DH@di.ku.dk} \\
\AND
Qiaoling Xiao \\
WUST-Madrid Complutense Institute \\
Wuhan, China \\
\texttt{CHRISTINA.XIAO@wust.edu.cn}
}

\iclrfinalcopy 
\begin{document}

\maketitle

\begin{abstract}
As Large Language Models (LLMs) increasingly influence high-stakes decision-making across global contexts, ensuring their alignment with diverse cultural values has become a critical governance challenge. This study presents a Multi-Layered Auditing Platform for Responsible AI that systematically evaluates cross-cultural value alignment in China-origin and Western-origin LLMs through four integrated methodologies: Ethical Dilemma Corpus for assessing temporal stability, Diversity-Enhanced Framework (DEF) for quantifying cultural fidelity, First-Token Probability Alignment for distributional accuracy, and Multi-stAge Reasoning frameworK (MARK) for interpretable decision-making. Our comparative analysis of 20+ leading models, such as Qwen, GPT-4o, Claude, LLaMA, and DeepSeek, reveals universal challenges—fundamental instability in value systems, systematic under-representation of younger demographics, and non-linear relationships between model scale and alignment quality—alongside divergent regional development trajectories. While China-origin models increasingly emphasize multilingual data integration for context-specific optimization, Western models demonstrate greater architectural experimentation but persistent U.S.-centric biases. Neither paradigm achieves robust cross-cultural generalization. We establish that Mistral-series architectures significantly outperform LLaMA-3-series in cross-cultural alignment, and that Full-Parameter Fine-Tuning on diverse datasets surpasses Reinforcement Learning from Human Feedback in preserving cultural variation. These findings provide empirical foundations for evidence-based AI governance, offering actionable protocols for model selection, bias mitigation, and policy consultation at scale, while demonstrating that current LLMs require sustained human oversight in ethical decision-making and cannot yet autonomously navigate complex moral dilemmas across cultural contexts.
\end{abstract}

\section{Introduction}

The integration of Large Language Models (LLMs) into high-stakes applications, such as decision-support systems, requires a thorough evaluation of their moral and ethical reasoning capabilities. For AI agents to operate trustworthily, their behavior must align with human values, which often vary significantly across different contexts and cultures~\citep{Liscio23Value}. Evaluating cultural value alignment is therefore a high priority within natural language processing and AI ethics research~\citep{jobin19global}.

Early work on machine ethics has focused on complex ethical dilemmas, which necessitate choices between two "right" options involving conflicting moral values~\citep{kidder1996good}. Comprehensive evaluations have revealed that both China and the West (e.g., Qwen2-72B and Claude-3.5-Sonnet) LLMs exhibit definitive preferences between major conflicting value pairs ~\citep{Yuan24Right}. Larger and more advanced LLMs tend to support a deontological perspective, maintaining their choices even when negative consequences are specified. However, LLMs often struggle with understanding the core decision-making task, demonstrating pronounced sensitivity to how dilemmas are formulated in prompts~\citep{Sclar24Quantifying, Yuan24Right}. Moreover, these models show significant cultural and linguistic biases, with moral reasoning varying substantially depending on the language used for prompting~\citep{Agarwal24Ethical}.

\begin{figure}
    \centering
    \includegraphics[width=\linewidth]{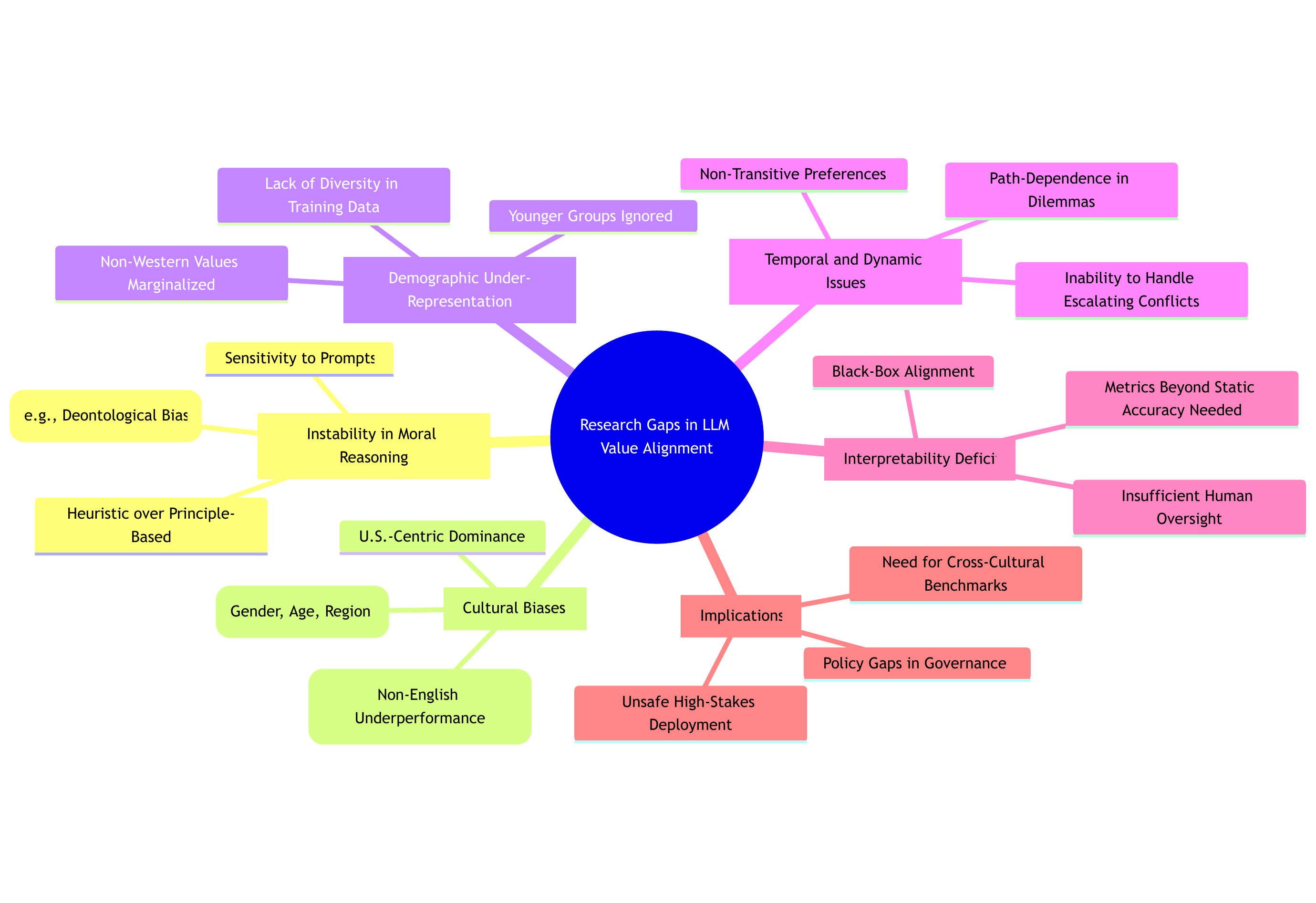}
    \caption{Key research gaps in cross-cultural value alignment for LLMs include: instability in moral reasoning (e.g., sensitivity to prompts and consequences), cultural biases (e.g., U.S.-centric or English-dominant preferences), under-representation of diverse demographics (e.g., younger groups or non-Western values), lack of temporal stability in ethical decisions, and insufficient interpretability in alignment methods.}
    \label{fig:gaps}
\end{figure}

These findings establish a critical governance constraint (Figure \ref{fig:gaps}): solely relying on LLMs to resolve contentious ethical issues on their own is neither safe nor desirable, nor is it ethical in critical applications, given the models' rigidity and tendency to favor certain values~\citep{Mittelstadt19Principles}. This volatility, combined with observed rigidity, underscores that simply providing instructions is insufficient for ensuring Responsible AI (RAI) deployment. A shift toward a dynamic, systematic auditing platform is necessary to measure LLM consistency, stability, fidelity, and bias, thereby advancing evaluation beyond static, single-step assessment methods~\citep{Scherrer23Evaluating}.

To address these issues, this paper systematically presents a \textbf{Multi-Layered Auditing Platform for Responsible AI}, composed of four integrated computational tools designed to assess LLM consistency, stability, fidelity, and interpretability in cross-cultural contexts: (1) \textbf{Ethical Dilemma Corpus}: moving beyond single-step assessments, the platform first diagnoses reliance on unstable heuristics and reveals value misalignment. (2) \textbf{Diversity-Enhanced Framework (DEF)}: utilized to overcome repetitive LLM output and generate accurate preference distributions reflecting the diversity and uncertainty inherent in survey responses. (3) \textbf{First-Token Probability Alignment technique}: fine-tunes LLMs to minimize divergence between predicted and actual human value distributions. (4) \textbf{Multi-stAge Reasoning frameworK (MARK)}: enhances accountability and interpretability by simulating human decisions through personality-driven cognitive processes, integrating stress analysis and multi-stage reasoning to provide psychologically grounded explanations.

The framework systematically integrates these methodologies to create a multi-layered platform capable of generating nuanced, cross-cultural insights that serve both academic transparency and strategic industrial value:

\begin{itemize}
    \item \textbf{Technology for Good (Social Impact)}: The methodologies provide safety guidance and protocols for investigating complex moral behaviors in LLMs. By enabling accurate simulation of group-level survey responses—a capability often limited by high costs and time-intensive human studies—these tools can accelerate social science research and aid in more informed policy decisions, particularly in navigating complex ethical choices~\citep{Lisa22Out, Bail21Can}.
    \item \textbf{Industrial Value (LLM Governance)}: The platform enables competitive strategic decisions by benchmarking LLMs across critical RAI dimensions, moving beyond generic accuracy metrics. This systematic audit identifies optimal model architectures and verifies the effectiveness of alignment techniques, providing a blueprint for building culturally sensitive and compliant global AI products~\citep{Hagendorff20The}.
\end{itemize}

This paper thus aims to bridge the gap between technical innovation and social/industrial utility by systematically identifying robust value alignment mechanisms for China-Western models.

\section{Related Works}
\subsection{AI Ethics and Computational Morality}
\paragraph{Moral Beliefs in LLMs}
\citet{Scherrer23Evaluating, durmus2024measuringrepresentationsubjectiveglobal} introduced a statistical method for eliciting beliefs encoded in LLMs through surveys of moral scenarios, distinguishing between high and low-ambiguity cases based on common morality rules. Their large-scale survey comprised 680 high-ambiguity and 687 low-ambiguity moral scenarios administered to 28 open- and closed-source LLMs, revealing that most models exhibit low uncertainty in unambiguous scenarios while expressing uncertainty in ambiguous cases.

\paragraph{Moral Foundations Theory} 
Recent work has applied Moral Foundations Theory (MFT) to measure how well LLMs represent different political orientations and ethical perspectives \citep{dev-etal-2022-measures, Narayanan25Ethical, Shaina24Nbias}. Studies employing experimental psychology methods to probe LLMs' moral and legal reasoning have found that alignment with human responses varies significantly across different experiments and models.

\paragraph{Ethical Dilemmas - Right vs. Right}
\citet{Yuan24Right} conducted a comprehensive evaluation using Kidder's framework to examine how LLMs navigate ethical dilemmas involving Truth vs. Loyalty, Individual vs. Community, Short-Term vs. Long-Term, and Justice vs. Mercy. Their study revealed that LLMs exhibit pronounced preferences, prioritizing truth over loyalty in 93.48\% of cases, long-term over short-term considerations in 83.69\% of cases, and community over individual in 72.37\% of cases. This work demonstrated that larger LLMs tend to support a deontological perspective, maintaining their choices even when negative consequences are specified.

\paragraph{Classical Ethical Theories}
\citet{agarwal-etal-2024-ethical} examined ethical reasoning across different languages, finding that GPT-4 is the most consistent ethical reasoner across languages, while other models show significant moral value bias when prompted in languages other than English. Their experiments covered deontology, virtue ethics, and consequentialism across six languages (English, Spanish, Russian, Chinese, Hindi, and Swahili). The AMULED framework translates utilitarianism, deontology, virtue ethics, and other moral philosophies into reward functions for reinforcement learning to address moral uncertainty in AI decision-making \citet{dubey2025addressingmoraluncertaintyusing}.

\subsection{LLM Value Alignment and Governance}
\paragraph{Moral Foundations Theory and Schwartz's Values}
\citet{Shoval24Assessing} applied Schwartz's Theory of Basic Values (STBV) to measure value-like constructs within LLMs, finding substantial divergence between LLM value profiles and population data. All models prioritized universalism and self-direction while de-emphasizing achievement, power, and security relative to humans. Their study using the Portrait Values Questionnaire-Revised (PVQ-RR) showed that these biased value profiles strongly predicted LLMs' responses when presented with mental health dilemmas requiring choosing between opposing values.

Recent work combining MFT and Schwartz's theory in multi-step moral dilemmas revealed that LLMs exhibit non-transitive and shifting moral preferences \citep{wu2025staircaseethicsprobingllm}. The study found that intuitive preferences like care decrease while fundamental values like fairness become more prominent as dilemmas progress, demonstrating that LLMs maintain value orientations while flexibly adjusting preference strengths across sequential dilemmas.

\paragraph{Value Alignment Methodologies}
Two primary approaches exist: data-driven bottom-up alignment, where LLM performance is evaluated by comparing moral judgment to human judgment using datasets like SOCIALCHEM101, Moral Stories, ETHICS, NormBank, and MoralChoice \citep{jiang2022machineslearnmoralitydelphi, emelin-etal-2021-moral, hendrycks2021ethics}; and top-down alignment, measuring how well LLMs infer specified value preferences through prompt embedding. \cite{Yuan24Right} showed that explicit guidelines are more effective in guiding LLMs' moral choices than in-context examples.

\paragraph{Dynamic and Temporal Moral Reasoning}
\citet{wu2025staircaseethicsprobingllm} introduced the Multi-step Moral Dilemmas (MMDs) framework, a path-dependent evaluation approach that captures temporal dynamics of moral judgment, addressing limitations of static single-step assessment methods. This research across 3,302 five-stage dilemmas revealed that LLMs maintain value orientations while flexibly adjusting preference strengths across sequential dilemmas, with value preferences shifting significantly as dilemmas progress.

\paragraph{RLHF vs. Fine-Tuning}
Reinforcement Learning from Human Feedback \citep[RLHF]{ouyang2022traininglanguagemodelsfollow} has become central in adapting AI models to human-centric expectations, particularly in the final stages of fine-tuning state-of-the-art models, though alignment can be biased by the group of humans providing feedback and may never satisfy everyone's preferences simultaneously \citep{christiano2023deepreinforcementlearninghuman, cao-etal-2025-specializing}. Studies on cultural value alignment found that Mistral-series models demonstrate superior performance compared to Llama-3-series models in cross-cultural value alignment contexts \citep{Liu25Towards}.

\subsection{LLM Simulation in Social Science and Psychology}
\paragraph{LLM Simulations of Human Behavior}
Researchers have pioneered specializing LLMs for simulating group-level survey response distributions for global populations, using fine-tuning methods based on first-token probabilities to minimize divergence between predicted and actual response distributions \citep{Argyle_Busby_Fulda_Gubler_Rytting_Wingate_2023}. The SocSci210 dataset, comprising 2.9 million responses from 400,491 participants across 210 social science experiments, has enabled LLMs to produce predictions 26\% more aligned with human response distributions.

\paragraph{Distribution Prediction and Calibration}
The task of simulating survey response distributions can be viewed as a calibration problem, aligning classifier predictive probabilities with classification uncertainty. Uncertainty quantification methods have been developed to convert simulated LLM responses into valid confidence sets for population parameters, addressing distribution shift between simulated and real populations \citep{pmlr-v202-santurkar23a}.

\paragraph{Psychologically Grounded Simulation}
The PsyDI framework incorporates MBTI psychological theory to model decision-making processes through progressively in-depth dialogue, leveraging cognitive functions to provide more accurate personality measurements. Fine-tuning approaches using sociodemographic prompting and persona strategies have been shown to enhance personalization in survey simulations \citep{Park23Generative, liu2025demographics}.

\subsection{Cross-Cultural Alignment, Bias, and Evaluation Metrics}
\paragraph{Cultural Alignment and Cross-Cultural NLP}
Evaluations using the World Values Survey have revealed that all major LLMs exhibit cultural values resembling English-speaking and Protestant European countries \citep{Tao24Cultural}. Cultural prompting improved alignment for 71-81\% of countries in later models like GPT-4o. Comprehensive evaluations across 10 models, 20 countries, and languages using Hofstede's Value Survey Module have identified systematic cultural biases favoring highly represented languages and regions \citep{cao-etal-2023-assessing}.

\paragraph{Evaluation of Cultural Biases and Stereotypes}
Research has shown that LLMs are 3-6 times more likely to choose occupations that stereotypically align with a person's gender, and these choices align more with people's perceptions than with ground truth job statistics \citep{kotek_gender_2023}. Studies using LLM Word Association Tests have found pervasive stereotype biases across 8 models in 4 social categories (race, gender, religion, health) covering 21 stereotypes, demonstrating that value-aligned models still harbor implicit biases \citep{taubenfeld-etal-2024-systematic}.

\paragraph{Cultural Datasets and Benchmarks} 
The Integrated Values Surveys (IVS), combining the World Values Survey and European Values Study, provide an established measure of cultural values for 112 countries and territories. The SOCIALCHEM101 dataset contains 292,000 sentences representing rules of thumb for evaluating LLMs' ability to reason about social and moral norms \citep{forbes-etal-2020-social}, while the Moral Integrity Corpus provides 38,000 prompt-reply pairs with 99,000 rules of thumb and 114,000 annotations.

\paragraph{Consistency and Bias Measurement}
Comprehensive taxonomies organize bias evaluation by metrics operating at different levels (embeddings, probabilities, generated text) and datasets by their structure, with techniques including counterfactual testing, stereotype detection, and sentiment analysis \citep{Navigli23Biases}. Diversity-Enhanced Frameworks have been developed to measure cultural value misalignment through multi-aspect evaluation metrics, revealing notable concerns regarding the lack of cross-cultural representation and preference biases related to gender and age.

\section{Methodologies: A Computational Auditing Platform for Responsible AI}

\begin{figure}
    \centering
    \includegraphics[width=0.8\linewidth]{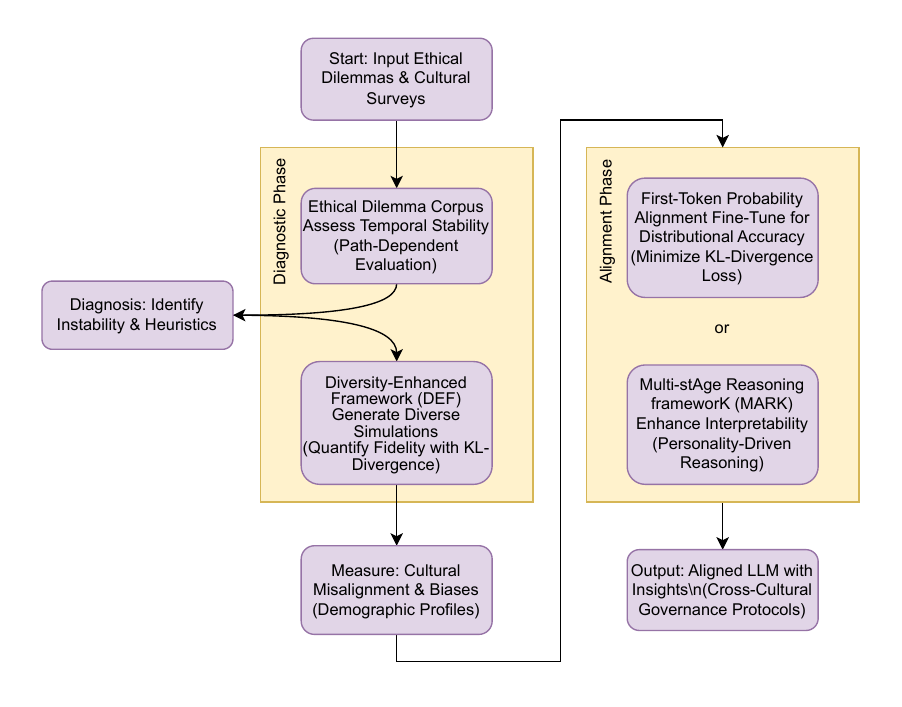}
    \caption{\textbf{Multi-Layered Auditing Platform for Responsible AI}, integrating four methodologies: Ethical Dilemma Corpus (for temporal stability), Diversity-Enhanced Framework (DEF, for cultural fidelity), First-Token Probability Alignment (for distributional accuracy), and Multi-stAge Reasoning frameworK (MARK, for interpretable decision-making).}
    \label{fig:pipe}
\end{figure}

In this section, we establish a systematic framework—a \textbf{Multi-Layered Auditing Platform for Responsible AI} (Figure \ref{fig:pipe})—which synthesizes and organizes existing computational tools derived from team member publications and relevant design efforts. This integrated approach moves beyond conventional, static evaluations of Large Language Models (LLMs) by addressing three fundamental research questions that guide our comparative analysis of \textbf{China-Western LLMs} in responsible AI governance and cross-cultural alignment.

Our framework systematically examines: (1) whether LLM responses to ethical dilemmas stem from stable value principles or contextual heuristics, (2) what quantifiable cultural values these models actually embody when compared to human populations, and (3) how current technological innovations can effectively contribute to genuine value alignment. Through this structured inquiry, we address the inherent rigidity and fixed value preferences (such as prioritizing \textbf{Truth over Loyalty} or adopting a \textbf{deontological perspective}) previously observed in LLMs when navigating ethical dilemmas. The core objective is to \textbf{audit LLM behaviors against specific Chinese and Western cultural human value distributions} to ensure genuine alignment, fairness, and safety in deployment.

\subsection{RQ1: Is Model Response to Ethical Dilemmas Rooted in Stable Value Principles?}

\begin{figure}
    \centering
    \includegraphics[width=0.7\linewidth]{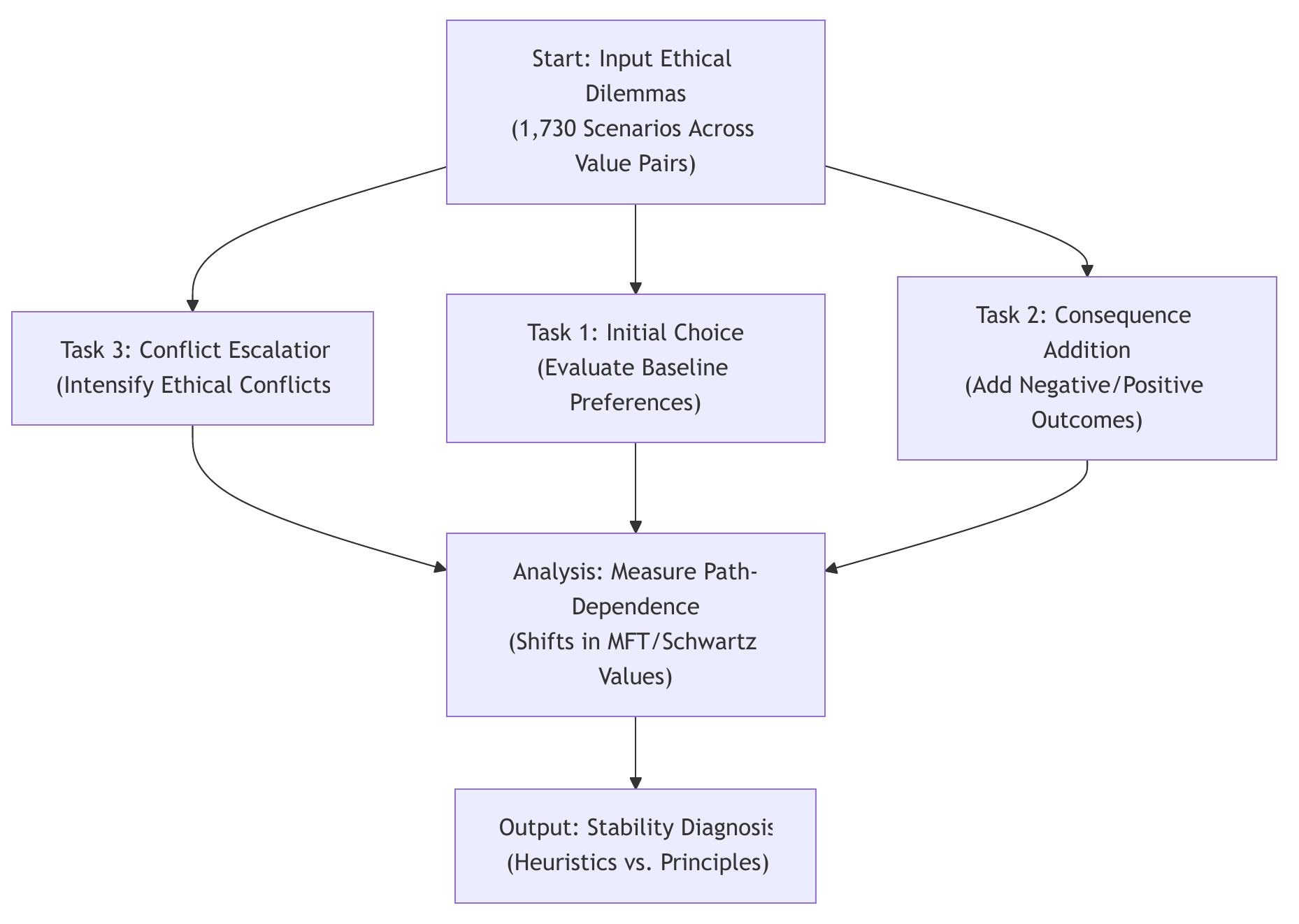}
    \caption{\textbf{Pipeline for Ethical Dilemma Corpus}: This component assesses temporal stability in LLM moral choices through path-dependent ethical dilemmas.}
    \label{fig:s1-pip}
\end{figure}

To investigate whether LLMs possess stable value principles or merely employ context-driven heuristics, we deploy the \textbf{Ethical Dilemma Corpus}~\citep{Yuan24Right} framework to audit the critical dimension of \textbf{Temporal Stability} (path-dependence) in LLM moral choices (Figure \ref{fig:s1-pip}).

Existing assessments primarily rely on single-step evaluations, which fail to capture how models adapt to evolving ethical challenges. The dataset addresses this gap by featuring 1,730 scenarios that progressively intensify ethical conflicts across three sequential tasks. This path-dependent evaluation framework enables a dynamic analysis of how LLMs adjust their moral reasoning when faced with escalating dilemmas.

The framework is applied to compare LLMs from both regions (including DeepSeek, LLaMA-3-70B, GPT-4o, and GLM-4) in terms of dynamic shifts in values (using Moral Foundations Theory or Schwartz's Theory of Basic Values) under conflict escalation.

\subsection{RQ2: What Quantifiable Cultural Values Do LLMs Obtain?}

\begin{figure}
    \centering
    \includegraphics[width=\linewidth]{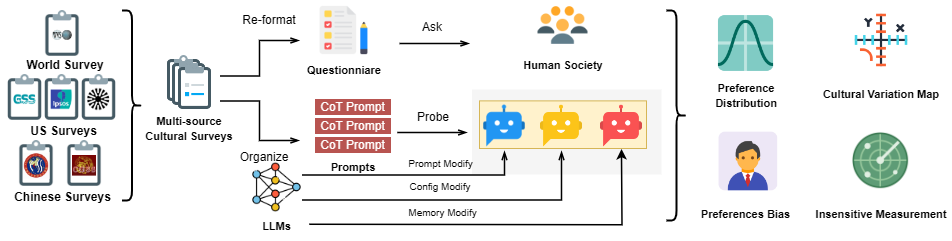}
    \caption{\textbf{Pipeline for Diversity-Enhanced Framework (DEF)}: DEF generates diverse simulations to quantify cultural fidelity against benchmarks like the World Values Survey. The pipeline focuses on overcoming repetitive outputs.  \textit{Adapted from \citet{Liu25Towards}.}}
    \label{fig:s2-pip}
\end{figure}

To audit Representativeness and \textbf{Cultural Fidelity} against external cultural benchmarks and quantify the cultural values embedded in LLMs, we utilize the \textbf{Diversity-Enhanced Framework}~\citep[DEF, Figure \ref{fig:s2-pip}]{Liu25Towards}. DEF systematically generates diverse virtual participants for high-fidelity simulation of human value survey data (e.g., the World Values Survey) across U.S. and Chinese cultures.

This framework is necessary because traditional LLM probing techniques, particularly those using Chain-of-Thought (CoT) instructions, often produce repetitive responses that fail to capture the necessary complexity and variability needed to generate accurate preference distributions. DEF captures the diversity and uncertainty inherent in LLM behaviors through realistic survey experiments by implementing:
\begin{itemize}
    \item \textbf{Prompt modifications} (e.g., randomly selected locations)
    \item \textbf{Configuration modifications} (e.g., dynamically tuning generation parameters like $num\_beams$)
    \item \textbf{Memory manipulation} (simulating memory effects and cleaning invalid responses)
\end{itemize}

The comparative assessment focuses on quantifying misalignment using metrics such as the Kullback-Leibler Divergence (KL-D) for \textbf{Preference Distribution} (to measure East-West divergence) and analyzing \textbf{Preference Bias} (demographic profiles) to expose fairness risks specific to the U.S. and Chinese contexts.

\subsection{RQ3: How Do Current Technological Innovations Contribute to Value Alignment?}
Having identified instability in value principles and measured cultural misalignment, we present two technological innovations that enhance value alignment in LLMs.

\subsubsection{First-Token Probability Alignment for Distributional Accuracy}
To transform the initial LLM response distributions toward high-fidelity representations of human data, the platform incorporates a specialized \textbf{First-Token Probability Alignment}~\citep{cao-etal-2025-specializing} technique. This Alignment Engineering solution fine-tunes models to minimize divergence against ground-truth human distributional preferences derived from surveys.

We devised this fine-tuning method based on first-token probabilities, where the objective is to minimize Kullback-Leibler Divergence loss between the predicted LLM distribution and the actual country-level human response distribution.

\subsubsection{Multi-stAge Reasoning frameworK (MARK) for Interpretable Alignment}
\begin{figure}
    \centering
    \includegraphics[width=\linewidth]{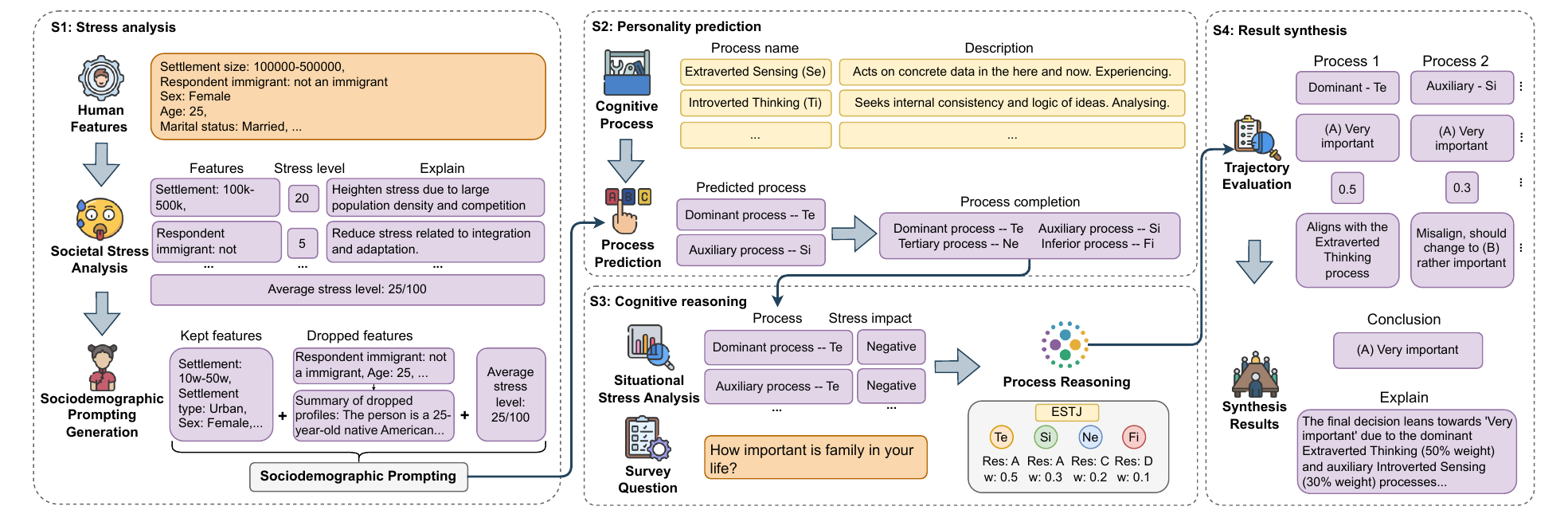}
    \caption{\textbf{Pipeline for Multi-stAge Reasoning frameworK (MARK)}: MARK enhances interpretability by simulating personality-driven reasoning based on MBTI theory. \textit{Adapted from \citet{liu2025demographics}.}}
    \label{fig:s4-pip}
\end{figure}

The \textbf{Multi-stAge Reasoning frameworK}~\citep[MARK, Figure \ref{fig:s4-pip}]{liu2025demographics} serves as a cognitive augmentation tool designed to enhance \textbf{Accountability and Interpretability} within the platform. MARK addresses the black-box nature of value alignment by simulating human decisions through integration of psychological theory—specifically the type dynamics theory in the MBTI psychological framework—to model cognitive processes and their interactions with stress and personal experience.

The framework systematically models human reasoning through multi-stage processing, including stress analysis, personality prediction, cognitive reasoning, and synthesis. This detailed simulation achieves high accuracy compared to other simulation baselines and demonstrates robust generalization using both \textbf{U.S. and Chinese survey data}. MARK provides \textbf{mechanistic, personality-driven explanations} for simulated value choices, enabling human experts to validate or challenge the model's reasoning trajectory. This computational interpretability is essential for building trust and ensuring accountability in cross-cultural AI applications.

Together, these components form a comprehensive auditing platform that not only diagnoses problems in current LLM value alignment (through Ethical Dilemma Corpus and DEF) but also provides actionable technological solutions (through First-Token Alignment and MARK). This integrated approach enables systematic comparison between Chinese and Western LLMs while advancing the technical capabilities needed for responsible cross-cultural AI governance.

\section{Cross-Cultural Audit Findings: Comparative Performance and Temporal Patterns in China-Western LLMs}

\begin{figure}
    \centering
    \includegraphics[width=0.85\linewidth]{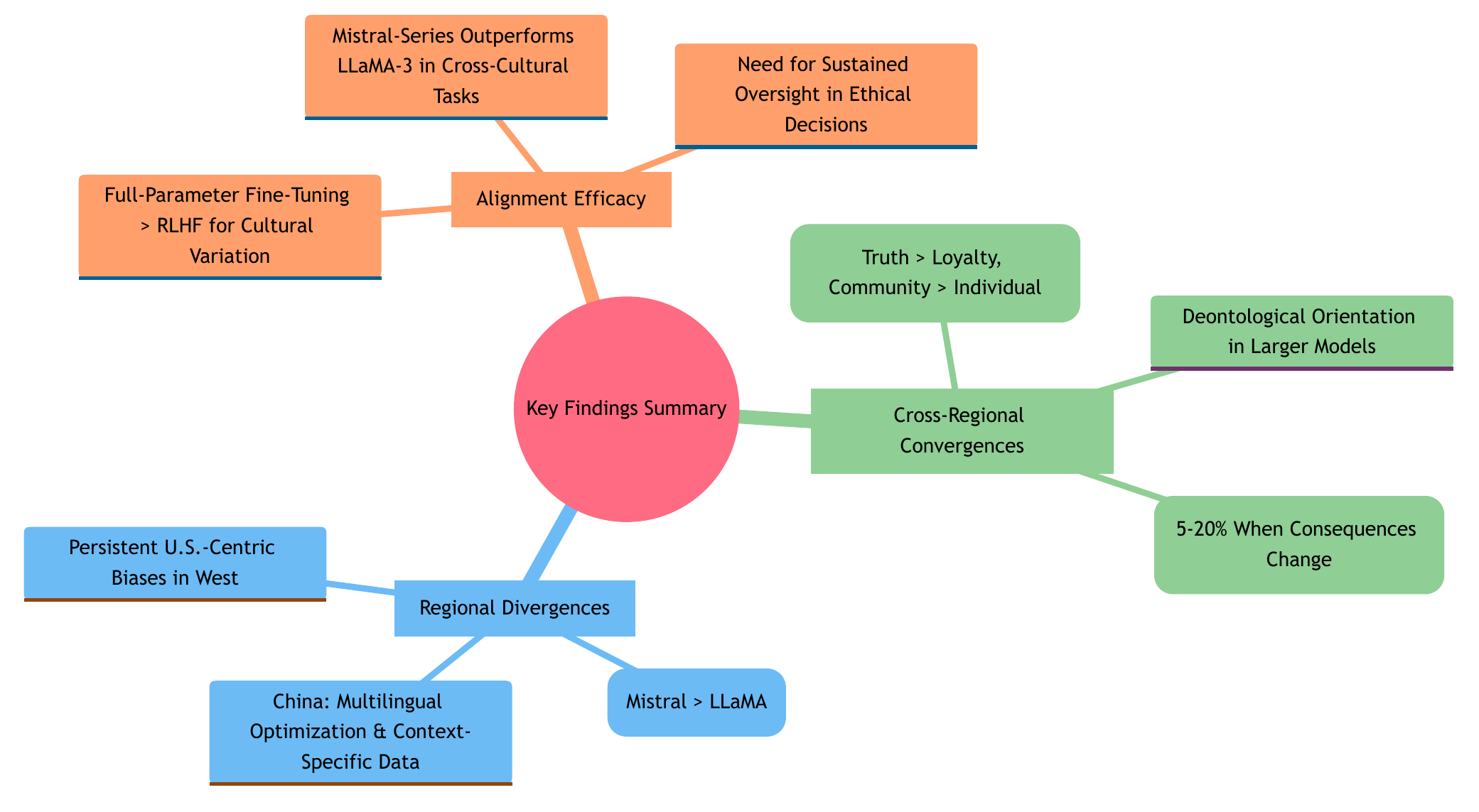}
    \caption{\textbf{Summary of Key Empirical Findings}: A mindmap overview of the cross-cultural audit findings, highlighting convergences (e.g., deontological trends) and divergences (e.g., optimization strategies).}
    \label{fig:find}
\end{figure}

\begin{figure}
    \centering
    \includegraphics[width=0.8\linewidth]{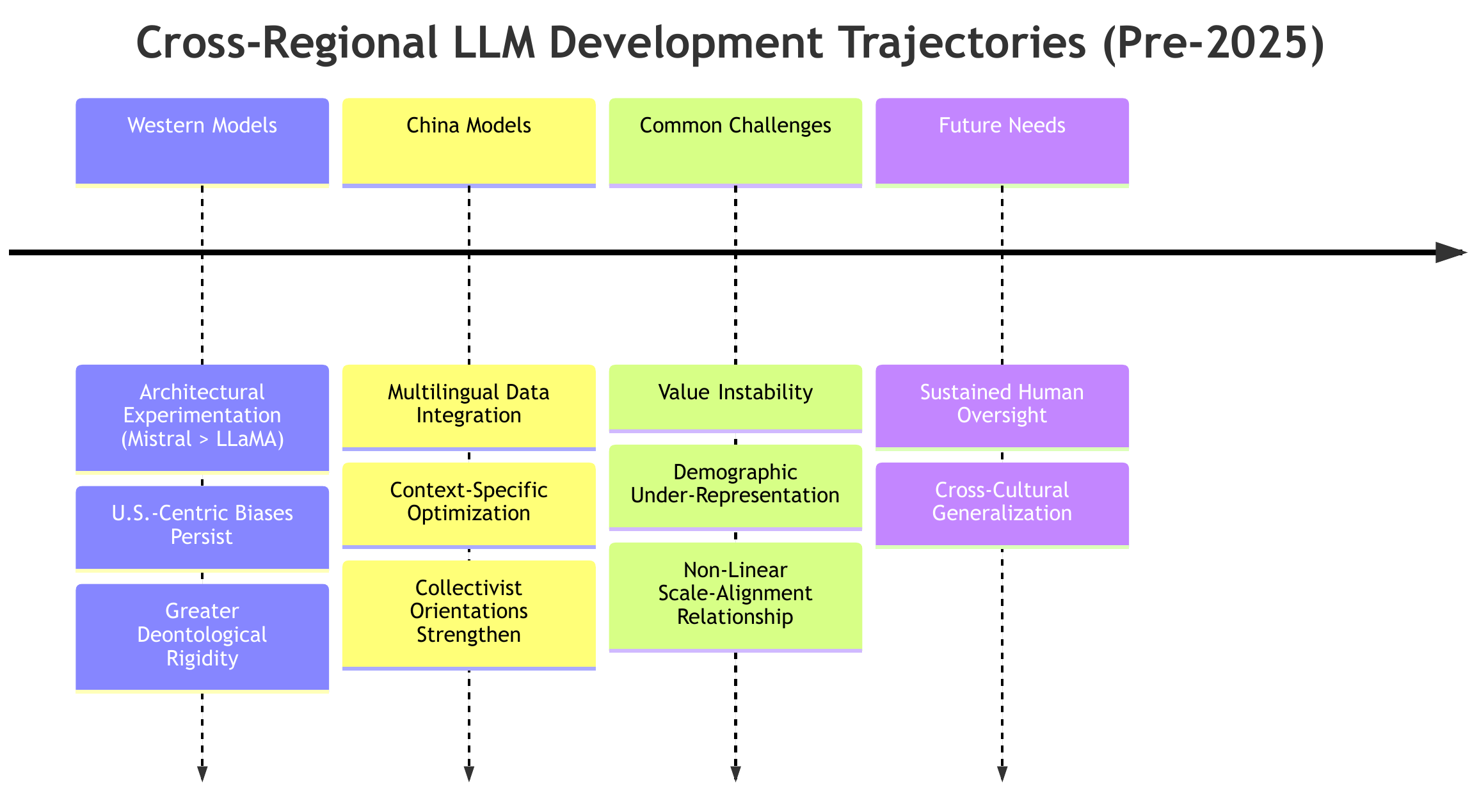}
    \caption{\textbf{Divergent China-West trajectories}: China focusing on multilingual data integration, West on architectural experimentation, but with persistent biases.}
    \label{fig:traj}
\end{figure}

This section systematically presents empirical findings derived from the proposed Multi-Layered Auditing Platform (summarized in Figure \ref{fig:find}), structured according to the three core research questions. The analysis emphasizes \textbf{comparative performance dynamics} between China-origin and Western-origin Large Language Models (LLMs), examining \textbf{cross-regional convergences and divergences} in stability, cultural fidelity, and alignment efficacy. Additionally, we investigate \textbf{temporal developmental patterns} (summarized in Figure \ref{fig:traj}) to assess whether technological advancement trajectories yield consistent improvements in value alignment across both regional contexts.

\subsection{Temporal Stability of Moral Reasoning (RQ1): Evaluating Value Principle Consistency in Ethical Decision-Making}

This subsection addresses RQ1 through the Ethical Dilemma Corpus \citep{Yuan24Right}, examining whether LLM responses to ethical dilemmas demonstrate stable, principle-based reasoning or context-dependent heuristic patterns. The analysis reveals both cross-regional architectural commonalities and region-specific developmental trajectories.

\subsubsection{Cross-Regional Convergence: Universal Value Rigidity and Deontological Orientation}

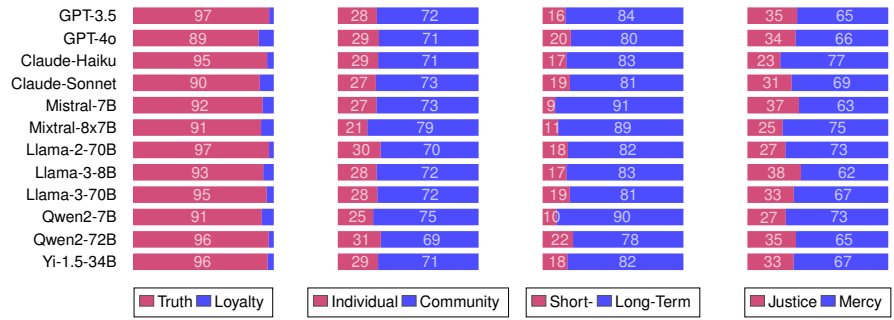
\begin{figure*}[]
\centering
\pgfplotsset{/pgfplots/bar cycle list/.style={/pgfplots/cycle list={
    }
}}%
\begin{tikzpicture}
\tikzstyle{every node}=[font=\footnotesize\sffamily,draw=none] 
\begin{groupplot}[
group style={columns=4, horizontal sep=25pt,}, 
height=7cm,
width=17cm]
\pgfplotsset{nodes near coords custom/.style={
            every node near coord/.style={
                check for small/.code={
                    \pgfkeys{/pgf/fpu=true}
                    \pgfmathparse{\pgfplotspointmeta<#1}%
                    \pgfkeys{/pgf/fpu=false}
                    \ifpgfmathfloatcomparison
                    \pgfkeysalso{above=0.5em}
                    \fi
                },
                check for small,
            },
        },
}
\pgfplotsset{compare axis style/.style={
    axis line style={opacity=0},
        xbar stacked,
        width=4.8cm, height=7cm,
        bar width=2.1ex,
        enlarge x limits={0.06},
        nodes near coords={
            \pgfkeys{/pgf/fpu=true}
            \pgfmathparse{\pgfplotspointmeta / 100 * 100}
            $\pgfmathprintnumber[fixed, precision=0]{\pgfmathresult}$
            \pgfkeys{/pgf/fpu=false}
        },
        nodes near coords custom=1,
        every node near coord/.append style={text=white,font=\footnotesize\sansmath\sffamily,/pgf/number format/.cd, precision=1,fixed,1000 sep={}},
        xmin=-1, xmax=101,
        xtick={0, 50, 100},
        xticklabels = {},
        tick style={draw=none},
        xtick pos=bottom,
        ytick=data,
        enlarge y limits=0.05,
        legend style={at={(0.5,-0.05)}, anchor=north, legend columns=-1},
}
}
\nextgroupplot[compare axis style,
        tick label style = {font=\footnotesize\sansmath\sffamily},
        table/col sep=comma,
        yticklabels from table={data/consistency.csv}{Model},
        ]
    \addplot[draw=none,fill=purple, opacity=0.7] table [col sep=comma,x=Truth,y expr=\coordindex] {data/consistency.csv};
    \addplot[draw=none,fill=blue, opacity=0.7,nodes near coords={}] table [col sep=comma,x=Loyalty,y expr=\coordindex] {data/consistency.csv};
    \legend{\strut Truth, \strut Loyalty}

\nextgroupplot[compare axis style,
            yticklabel=\empty,
        ]
    \addplot[draw=none,fill=purple, opacity=0.7] table [col sep=comma,x=Individual,y expr=\coordindex] {data/consistency.csv};
    \addplot[draw=none,fill=blue, opacity=0.7] table [col sep=comma,x=Community,y expr=\coordindex] {data/consistency.csv};
    \legend{\strut Individual, \strut Community}
 
\nextgroupplot[compare axis style,
            yticklabel=\empty,
        ]
    \addplot[draw=none,fill=purple, opacity=0.7] table [col sep=comma,x=Short-Term,y expr=\coordindex] {data/consistency.csv};
    \addplot[draw=none,fill=blue, opacity=0.7] table [col sep=comma,x=Long-Term,y expr=\coordindex] {data/consistency.csv};
    \legend{\strut Short-, \strut Long-Term}

\nextgroupplot[compare axis style,
            yticklabel=\empty,
        ]
    \addplot[draw=none,fill=purple, opacity=0.7] table [col sep=comma,x=Justice,y expr=\coordindex] {data/consistency.csv};
    \addplot[draw=none,fill=blue, opacity=0.7] table [col sep=comma,x=Mercy,y expr=\coordindex] {data/consistency.csv};
    \legend{\strut Justice, \strut Mercy}
\end{groupplot}
\end{tikzpicture}
\caption{Moral value preference query results for each conflicting value pair. \textit{Reprinted from Jiaqing Yuan, et al. Right vs. right: Can llms make tough choices? CoRR, abs/2412.19926, 2024. doi: 10.48550/ARXIV.2412.1992, Copyright 2024.}}
\label{figure:choice}
\end{figure*}

\paragraph{Comparative baseline performance} As presented in figure \ref{figure:choice}, LLMs from both China and Western development ecosystems exhibit remarkably consistent preferential hierarchies when navigating "Right vs. Right" ethical dilemmas involving conflicting moral values, with pronounced preferences for Truth over Loyalty, Long-term over Short-term considerations, and Community over Individual interests, suggesting convergent training paradigms despite distinct cultural contexts and data sources. The Individual-Community value pair demonstrates the highest rigidity across all models, indicating deeply embedded collectivist orientations that transcend regional boundaries.

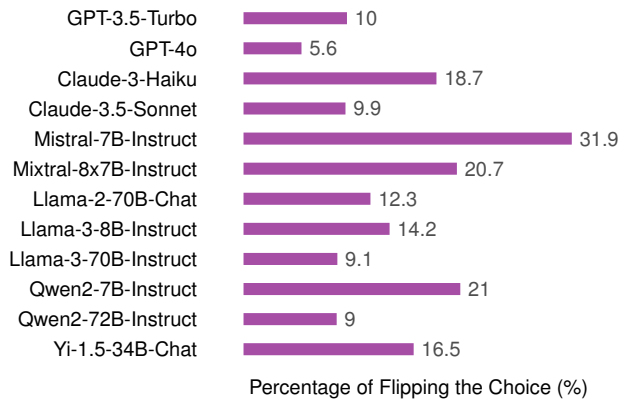
\begin{figure}[]
\centering
\pgfplotsset{/pgfplots/bar cycle list/.style={/pgfplots/cycle list={
    {draw=none, opacity=0.7, fill=violet},
    {draw=none, opacity=0.7, fill=teal},
    {draw=none, opacity=0.7, fill=purple},
    {draw=none, opacity=0.7, fill=blue},
    {draw=none, opacity=0.7, fill=olive},
    {draw=none, opacity=0.7, fill=magenta},
    {draw=none, opacity=0.7, fill=cyan},
    }
}}%
\begin{tikzpicture}
\tikzstyle{every node}=[font=\footnotesize\sffamily]
\begin{axis}[
axis line style={opacity=0},
    xbar=1pt,
    xtick={20,40,...,80}, 
    xticklabels= \empty,
    nodes near coords,
    every node near coord/.append style={font=\footnotesize\sansmath\sffamily,/pgf/number format/.cd, precision=1,fixed,1000 sep={}},
    enlarge x limits=0.05,
    enlarge y limits=0.02,
    xmin=-1,xmax=35,
    tick style={draw=none},
    tick label style = {font=\footnotesize\sansmath\sffamily},
    ytick=data,
    table/col sep=comma,
    yticklabels from table={data/consequence-flip.csv}{Model},
    height=7cm,
    width=8cm,
    bar width=1.2ex,
    xlabel={Percentage of Flipping the Choice (\%)},
    legend style={at={(0.2,-0.05)}, anchor=north,legend columns=1},
    legend cell align={left},
    legend image code/.code={\draw[draw=none,#1] (0cm,-0.1cm) rectangle (0.6cm,0.1cm);},
]

\addplot table [col sep=comma,x=Flipping-ratio,y expr=\coordindex] {data/consequence-flip.csv};

\end{axis}
\end{tikzpicture}
\caption{Percentage of flipping the choice when consequences are altered, e.g., an LLM switches from ``Action A'' to ``Action B'' when a negative outcome is added to ``Action A'' and a positive outcome is added to ``Action B''. \textit{Reprinted from Jiaqing Yuan, et al. Right vs. right: Can llms make tough choices? CoRR, abs/2412.19926, 2024. doi: 10.48550/ARXIV.2412.1992, Copyright 2024.}}
\label{figure:consequence-flip}
\end{figure}

\paragraph{Deontological convergence in advanced models} Both China-origin and Western-origin flagship models converge toward deontological ethical frameworks as model capacity increases. These advanced systems maintain initial moral commitments despite explicitly specified negative consequences (Figure \ref{figure:consequence-flip}), demonstrating outcome-independent reasoning. Advanced models from both regional contexts demonstrate comparably strong commitment stability, with flagship models from both China (e.g., Qwen2-72B) and Western contexts (e.g., GPT-4o, Llama-3-70B) showing similarly low decision reversal rates when confronted with consequence specifications, suggesting that architectural scale rather than regional origin determines deontological adherence.

\subsubsection{Regional Divergence: Volatility Patterns and Developmental Trajectories}
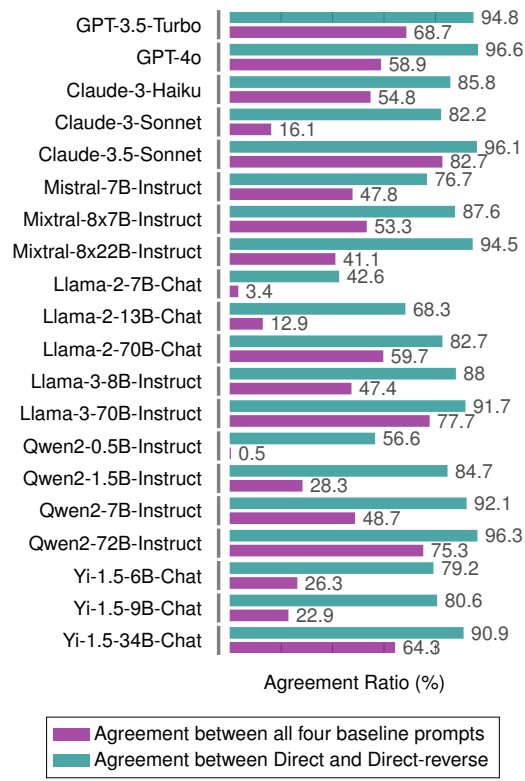
\begin{figure}[]
\centering
\pgfplotsset{/pgfplots/bar cycle list/.style={/pgfplots/cycle list={
    {draw=none, opacity=0.7, fill=violet},
    {draw=none, opacity=0.7, fill=teal},
    {draw=none, opacity=0.7, fill=purple},
    {draw=none, opacity=0.7, fill=blue},
    {draw=none, opacity=0.7, fill=olive},
    {draw=none, opacity=0.7, fill=magenta},
    {draw=none, opacity=0.7, fill=cyan},
    }
}}%
\begin{tikzpicture}
\tikzstyle{every node}=[font=\footnotesize\sffamily]
\begin{axis}[
axis line style={opacity=0},
after end axis/.code={
\foreach \y in {0,1,...,19}
    \draw [ultra thick, gray] ({rel axis cs:0,0}|-{axis cs:0,\y})
          ++(0pt,-1.25*\pgfkeysvalueof{/pgf/bar width})
          -- ({rel axis cs:0,0}|-{axis cs:0,\y})
          -- ++(0pt,1.25*\pgfkeysvalueof{/pgf/bar width});
  },
    xbar=1.5pt,
    xtick={20,40,...,80}, 
    xticklabels = {},
    nodes near coords,
    every node near coord/.append style={font=\footnotesize\sansmath\sffamily,/pgf/number format/.cd, precision=1,fixed,1000 sep={}},
    enlarge x limits=0.05,
    enlarge y limits=0.02,
    ytick style={draw=none},
    tick label style = {font=\footnotesize\sansmath\sffamily},
    ytick=data,
    table/col sep=comma,
    yticklabels from table={data/agreements.csv}{Model},
    height=12cm,
    width=6cm,
    bar width=1.2ex,
    xlabel={Agreement Ratio (\%)},
    legend style={at={(0.2,-0.1)}, anchor=north,legend columns=1},
    legend cell align={left},
    legend image code/.code={\draw[draw=none,#1] (0cm,-0.1cm) rectangle (0.6cm,0.1cm);},
]

\addplot table [col sep=comma,x=Agreement-all,y expr=\coordindex] {data/agreements.csv};

\addplot table [col sep=comma,x=Agreement-basic,y expr=\coordindex] {data/agreements.csv};

\legend{Agreement between all four baseline prompts,
Agreement between Direct and Direct-reverse};
\end{axis}
\end{tikzpicture}
\caption{Moral choice agreement for different prompts. The higher the agreement, the better the task comprehension. \textit{Reprinted from Jiaqing Yuan, et al. Right vs. right: Can llms make tough choices? CoRR, abs/2412.19926, 2024. doi: 10.48550/ARXIV.2412.1992, Copyright 2024, with permission from arXiv.}} 
\label{figure:agreement}
\end{figure}

\paragraph{Temporal improvements in task comprehension} Longitudinal analysis across model generations reveals consistent improvement in agreement ratios (task comprehension metrics) for both regional development tracks. Recent flagship releases from both China-origin models (e.g., Qwen2-72B) and Western-origin models (e.g., Claude-3.5-Sonnet, Llama-3-70B) achieve high agreement ratios across prompt variations, suggesting that increased training compute and data quality, rather than region-specific methodologies, drive comprehension improvements.

\paragraph{Fundamental instability in value systems} Despite surface-level improvements, the core finding transcends regional boundaries: LLM value systems remain fundamentally heuristic and unstable, relying on context-driven statistical pattern matching rather than globally consistent ethical principles. However, volatility patterns manifest distinctly across regional model families:
\begin{itemize}
    \item \textit{Western models}: \textbf{Claude} series demonstrates exceptional cross-context stability with minimal rank volatility
    \item \textit{China models}: \textbf{GLM-4-Air} exhibits comparable temporal consistency
    \item \textit{Volatile Western models}: \textbf{Llama} series shows pronounced rank fluctuations across value dimensions
    \item \textit{Volatile China models}: \textbf{DeepSeek} series demonstrates similar instability patterns
\end{itemize}

\paragraph{Sequential patterns across dilemma escalation} Analysis of the five-step Ethical Dilemma Corpus progression reveals universal and divergent patterns:
\begin{itemize}
    \item \textbf{Universal stability anchor}: The \textbf{Care} dimension maintains exceptional consistency across all escalation stages for both China and Western models, functioning as a stable moral foundation regardless of regional origin or training paradigm.
    \item \textbf{Progressive divergence in Authority}: The \textbf{Authority} dimension exhibits systematic degradation in cross-model consistency as dilemma severity increases. This pattern suggests that hierarchical value orientations, potentially influenced by distinct cultural training data, become increasingly model-specific under ethical pressure.
\end{itemize}

\subsection{Quantification of Cultural Value Encoding (RQ2): Cross-Regional Distributional Fidelity and Systemic Bias Patterns}
This subsection employs the Diversity-Enhanced Framework (DEF) \citep{Liu25Towards} to quantify cross-cultural misalignment and demographic preference bias, providing systematic architectural comparisons between China-origin and Western-origin foundational models.

\subsubsection{Architectural Performance: Cross-Regional Model Family Comparisons}

\begin{table}[]
\caption{The statistical result of t-test validation between Mistral- and Llama-3-series models. The result suggests that the Mistral-series models generally have significantly better value alignment performance than the Llama-3-series models. \textit{Adapted from \citet{Liu25Towards}.}}
\label{tbl:t_tst}
\begin{tabular}{l|lccc|cccc}
\toprule
\textbf{Culture}                  & \textbf{Model}          & \textbf{Mean} & \textbf{SD} & \textbf{SEM} & \textbf{t value}      & \textbf{DF}         & \textbf{SED}          & \textbf{p value}        \\ \midrule
\multirow{2}{*}{\textbf{U.S.}}     & \textbf{Llama-3-series} & 1.28          & 0.50        & 0.10         & \multirow{2}{*}{3.01} & \multirow{2}{*}{52} & \multirow{2}{*}{0.12} & \multirow{2}{*}{0.0041} \\
                                  & \textbf{Mistral-series} & 0.93          & 0.34        & 0.07         &                       &                     &                       &                         \\ \midrule
\multirow{2}{*}{\textbf{Chinese}} & \textbf{Llama-3-series} & 1.35          & 0.74        & 0.14         & \multirow{2}{*}{2.77} & \multirow{2}{*}{52} & \multirow{2}{*}{0.16} & \multirow{2}{*}{0.0079} \\
                                  & \textbf{Mistral-series} & 0.92          & 0.33        & 0.06         &                       &                     &                       &                         \\ \bottomrule
\end{tabular}
\end{table}

\paragraph{Western architectural comparison} Statistical analysis (Table \ref{tbl:t_tst}) of Western foundational architectures reveals significant performance stratification. The Mistral-series demonstrates substantially superior cultural value alignment compared to the Llama-3-series across both U.S. and Chinese cultural contexts (p = 0.0041 for U.S.; p = 0.0079 for Chinese, Wilcoxon signed-rank test). This finding suggests that the Mistral architecture encodes more \textbf{generalizable cultural representation capacities}, potentially attributable to training data diversity or architectural inductive biases that better capture cross-cultural value distributions.

\begin{figure}[]
    \centering
    \includegraphics[width=\textwidth]{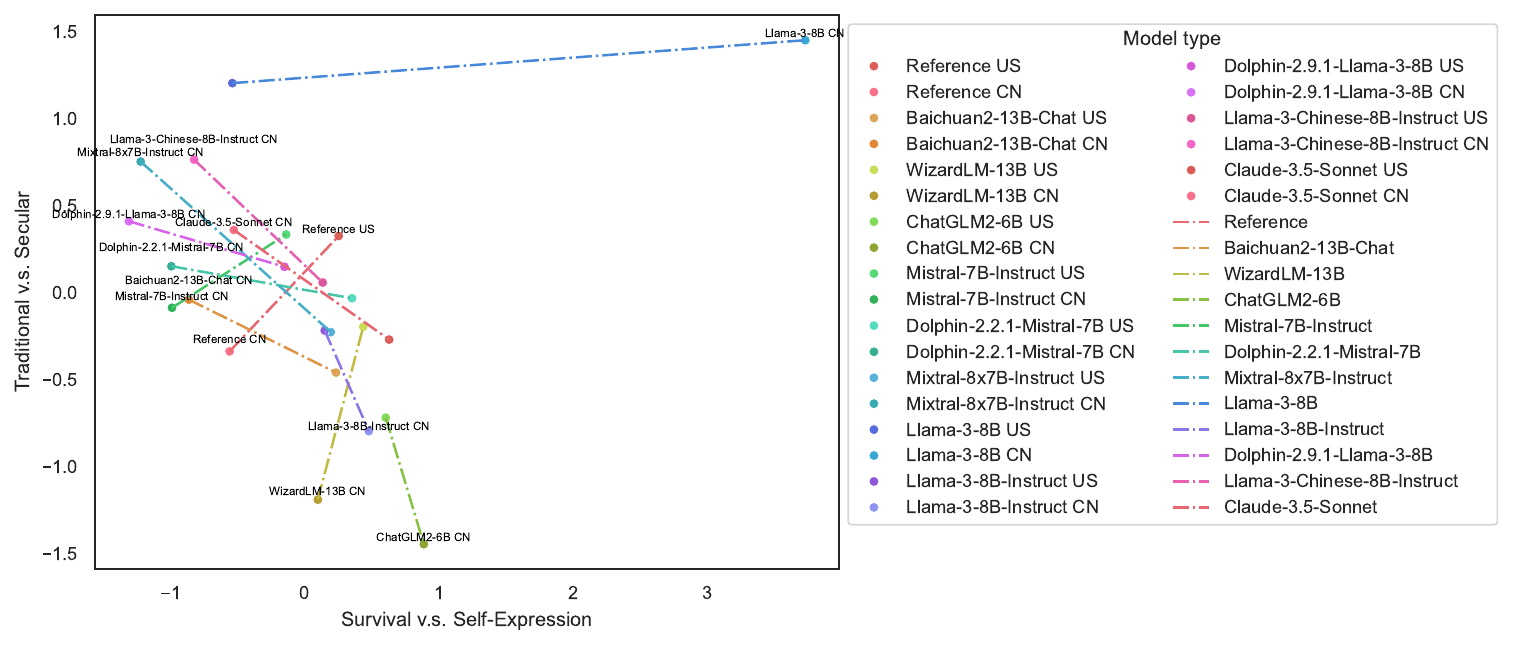}
    \caption{The cross-cultural variation map of model candidates. All models can somehow distinguish the cultural variations. Mistral-7B-Instruct preserves better cross-cultural variations. \textit{Adapted from \citet{Liu25Towards}.}}
    \label{fig:cumap}
\end{figure}

\paragraph{Cultural affinity patterns} While all evaluated models successfully distinguish between U.S. and Chinese value distributions on the Cultural Variation Map, systematic biases emerge:

For \textit{Western model biases}: \textbf{Llama-3-8B-Instruct} exhibits pronounced U.S. cultural affinity. Most Western models demonstrate U.S.-centric value alignment, with the notable exception of base \textbf{Llama-3-8B}, which shows reduced directional bias.

For the \textit{China model} performance: \textbf{Llama-3-Chinese-8B-Instruct} (Western architecture with enhanced Chinese training) demonstrates strong Chinese cultural affinity. \textbf{ChatGLM2-6B} uniquely balances U.S. and Chinese value representations among China-origin models

\paragraph{Cross-cultural preservation} \textbf{Mistral-7B-Instruct} achieves superior preservation of cultural variation across both contexts, outperforming models that achieve high alignment in single-culture scenarios, suggesting distinct optimization targets in Mistral's training regime.

\subsubsection{Systemic Failures and Demographic Bias: China-Western Contrast}

\begin{figure}[]
    \centering
    \subfloat[US survey]{
    \includegraphics[width=.33\linewidth]{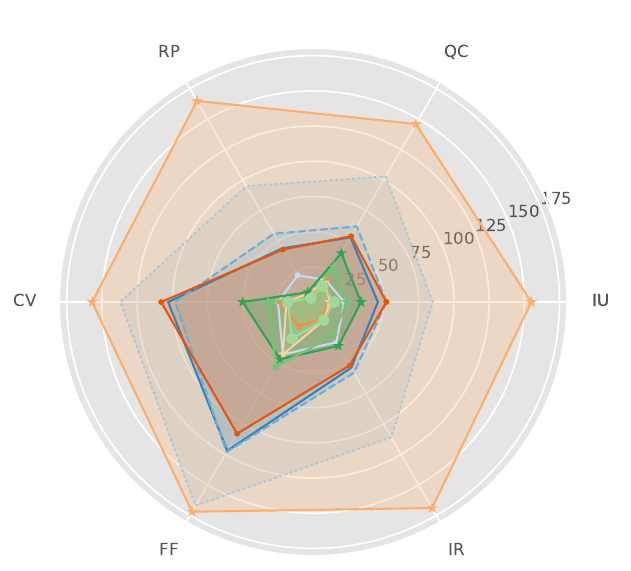}
    }
    \subfloat[CN survey]{
    \includegraphics[width=.52\linewidth]{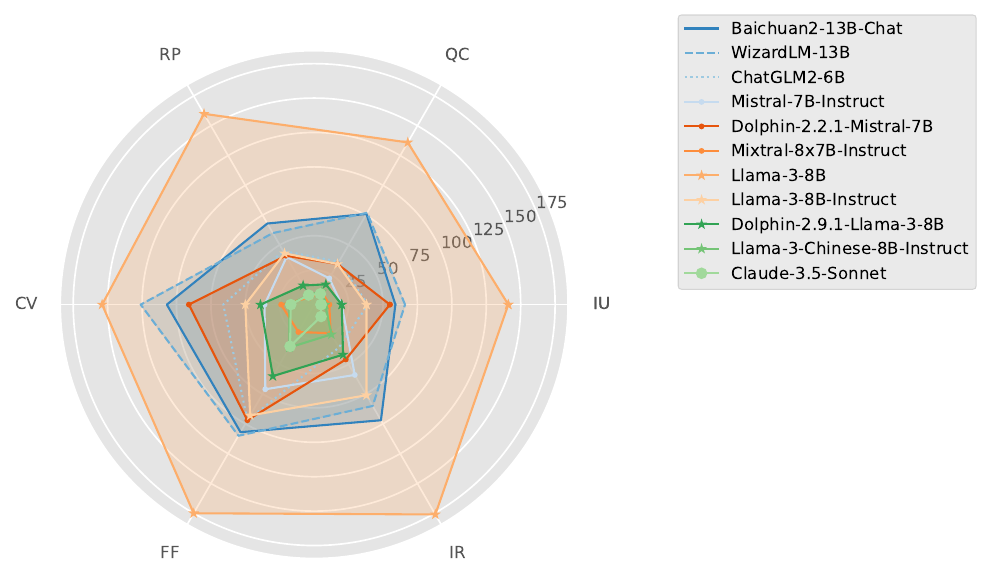}
    }
    \caption{The overall comparison of the insensitivity measurement among models on each survey. Model shows significant issues with \ff\ and \cv, with larger models having fewer problems on the US survey, and smaller models on the CN survey. \textit{Adapted from \citet{Liu25Towards}.}}
    \label{fig:ins-overall}
\end{figure}

\paragraph{Regional patterns in systemic inconsistency} Analysis of insensitivity measurements (Figure \ref{fig:ins-overall})—\textbf{False Fact Presentation (FF)} and \textbf{Conflict in Value Expression (CV)}—reveals consistent challenges across all models, but with significant architectural variation. The \textbf{Llama-3-series} exhibits approximately \textbf{2× the failure rate} of the Mistral-series across both cultural contexts, indicating fundamental differences in consistency mechanisms independent of regional adaptation efforts.

\paragraph{Demographic preference bias divergence} The audit reveals systematically distinct demographic mismatch profiles between cultural contexts (see Appendix \ref{app:bias}).

In the context of the United States, the observed bias pattern in alignment configurations demonstrates a preference for male demographic profiles. Additionally, there is a notable inclination towards the age groups of 30 to 49 years and 50 years and older, while individuals under the age of 29 are systematically under-represented.

In the context of Chinese demographics, the observed bias pattern demonstrates a preference for female profiles in optimal alignment configurations. There is a pronounced concentration within the age range of 30 to 49 years, accompanied by a systematic under-representation of persons younger than 29 years.

These divergent patterns suggest that training data demographic distributions, rather than intentional design choices, drive regional bias profiles. The universal under-representation of younger demographics across both contexts indicates a \textbf{shared data collection challenge} in both regional AI development ecosystems.

\subsection{Technological Innovation and Alignment Efficacy (RQ3): Developmental Trajectories and Cross-Regional Methodological Comparison
}
This subsection examines how technological development stages and innovation strategies influence value alignment, analyzing both \textbf{temporal progression patterns} and \textbf{methodological divergence} between China and Western AI development paradigms.

\subsubsection{Generational Scaling Effects: Non-Linear Relationships Between Capacity and Alignment}
\paragraph{Temporal trends in baseline capabilities} Longitudinal analysis across model generations confirms consistent improvements in task comprehension and response stability with increased model capacity and training sophistication. However, cultural alignment quality and demographic representation fidelity demonstrate non-monotonic relationships with scale, challenging assumptions of linear improvement trajectories.

\paragraph{The scale-alignment paradox} Comparative analysis of the Mixture-of-Experts (MoE) architecture Mixtral-8x7B-Instruct (Western) versus smaller dense models reveals critical tradeoffs:
\begin{itemize}
    \item \textit{Consistency advantages}: Mixtral-8x7B-Instruct exhibits significantly reduced insensitivity errors (FF and CV), indicating superior internal coherence
    \item \textit{Representation limitations}: Despite improved consistency, the larger MoE model demonstrates different patterns in cultural representation compared to the smaller \textbf{Mistral-7B-Instruct}
\end{itemize}

This finding suggests that standard scaling approaches, absent targeted alignment interventions, may inadvertently affect representation patterns while improving surface-level consistency.

\begin{table*}[!]
\centering
\resizebox{0.86\textwidth}{!}{
\begin{tabular}{l|l|cccccc}
\toprule
\multirow{2}{*}{Model} &
  \multirow{2}{*}{Methods} &
  \multicolumn{6}{c}{($1-$JSD) $\uparrow$} \\ \cmidrule{3-8} 
 &
   &
  $C_1$-$Q_3$ &
  $C_2$-$Q_1$ &
  $C_2$-$Q_3$ &
  $C_3$-$Q_1$ &
  \multicolumn{1}{c|}{$C_3$-$Q_3$} &
  \textit{Avg.} \\ \midrule
\multirow{4}{*}{\textit{Llama3-8B-Base}} &
  ZS [ctrl] &
  0.748 &
  0.766 &
  0.757 &
  0.779 &
  \multicolumn{1}{c|}{0.768} &
  0.764 \\
 &
  ZS &
  0.749 &
  0.768 &
  0.759 &
  0.781 &
  \multicolumn{1}{c|}{0.770} &
  0.765 \\
 &
  FT [ctrl] &
  0.756 &
  0.823 &
  0.751 &
  0.837 &
  \multicolumn{1}{c|}{0.770} &
  0.787 \\
 &
  \textbf{FT} &
  \textbf{0.770} &
  \textbf{0.858} &
  \textbf{0.773} &
  \textbf{0.877} &
  \multicolumn{1}{c|}{\textbf{0.781}} &
  \textbf{0.812} \\ \midrule
\multirow{4}{*}{\textit{Llama3-8B-Instruct}} &
  ZS [ctrl] &
  0.574 &
  0.626 &
  0.563 &
  0.644 &
  \multicolumn{1}{c|}{0.587} &
  0.599 \\
 &
  ZS &
  0.585 &
  0.650 &
  0.589 &
  0.657 &
  \multicolumn{1}{c|}{0.584} &
  0.613 \\
 &
  FT [ctrl] &
  0.756 &
  0.826 &
  0.751 &
  0.833 &
  \multicolumn{1}{c|}{0.764} &
  0.786 \\
 &
  \textbf{FT} &
  \textbf{0.777} &
  \textbf{0.881} &
  \textbf{0.783} &
  \textbf{0.890} &
  \multicolumn{1}{c|}{\textbf{0.784}} &
  \textbf{0.823} \\ \midrule
\multirow{4}{*}{\textit{Distil-Qwen-7B}} &
  ZS[ctrl] &
  0.586 &
  0.641 &
  0.642 &
  0.698 &
  \multicolumn{1}{c|}{0.682} &
  0.650 \\
 &
  ZS &
  0.583 &
  0.645 &
  0.639 &
  0.701 &
  \multicolumn{1}{c|}{0.671} &
  0.648 \\
 &
  FT[ctrl] &
  0.747 &
  0.764 &
  0.791 &
  0.811 &
  \multicolumn{1}{c|}{0.817} &
  0.786 \\
 &
  \textbf{FT} &
  \textbf{0.756} &
  \textbf{0.781} &
  \textbf{0.803} &
  \textbf{0.833} &
  \multicolumn{1}{c|}{\textbf{0.834}} &
  \textbf{0.801} \\ \midrule
\multirow{4}{*}{\textit{Distil-Qwen-14B}} &
  ZS[ctrl] &
  0.650 &
  0.704 &
  0.749 &
  0.711 &
  \multicolumn{1}{c|}{0.743} &
  0.712 \\
 &
  ZS &
  0.654 &
  0.716 &
  0.756 &
  0.731 &
  \multicolumn{1}{c|}{0.746} &
  0.721 \\
 &
  FT[ctrl] &
  0.751 &
  0.784 &
  0.799 &
  0.807 &
  \multicolumn{1}{c|}{0.833} &
  0.795 \\
 &
  \textbf{FT} &
  \textbf{0.777} &
  \textbf{0.816} &
  \textbf{0.827} &
  \textbf{0.851} &
  \multicolumn{1}{c|}{\textbf{0.861}} &
  \textbf{0.826} \\ \bottomrule
\end{tabular}
}
\caption{\label{tb:main_results}\textbf{Main results for predicting country-level survey response distributions on the WVS data}.
We test all models with zero-shot prompting (ZS) and our proposed fine-tuning approach (FT). [ctrl] indicates a control setup, where we randomly replace countries in test prompts with other countries, to evaluate country context sensitivity.
We report Jensen-Shannon Divergence ($1-$JSD, $\uparrow$). \textit{Adapted from \citet{cao-etal-2025-specializing}.}}
\end{table*}

\paragraph{Specialized alignment efficacy} As illustrated in Table \ref{tb:main_results}, the First-Token Alignment technique demonstrates the viability of targeted architectural interventions by using fine-tuning based on first-token probabilities to minimize divergence between predicted and actual response distributions, achieving substantial accuracy improvements over zero-shot baselines, confirming that \textbf{post-hoc alignment engineering} can effectively address distributional inaccuracies introduced during pretraining, regardless of model regional origin.

\subsubsection{Methodological Divergence: Comparative Efficacy of Alignment Paradigms}

\begin{figure*}[t]
    \centering
    \subfloat[US survey]{
        \includegraphics[width=0.42\textwidth]{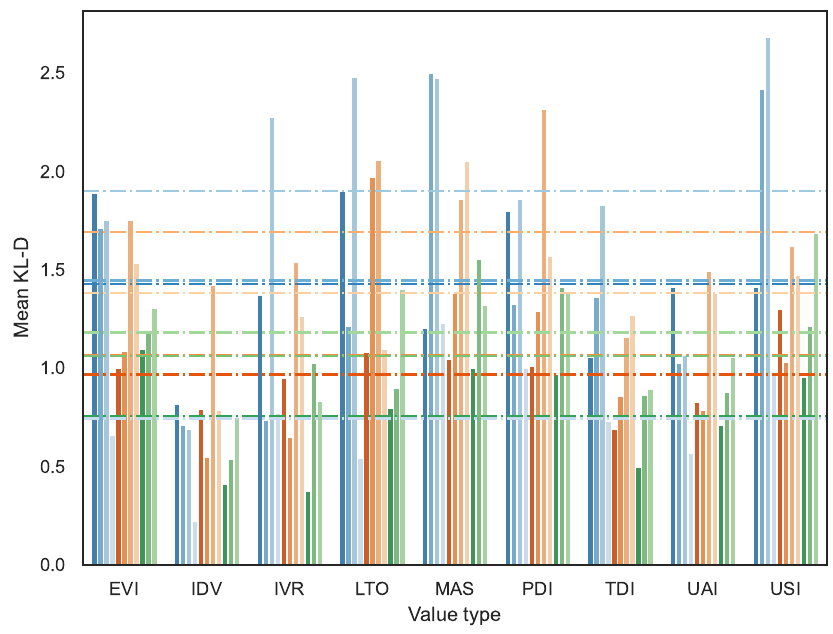}
    }
    \subfloat[CN survey]{
        \includegraphics[width=0.575\textwidth]{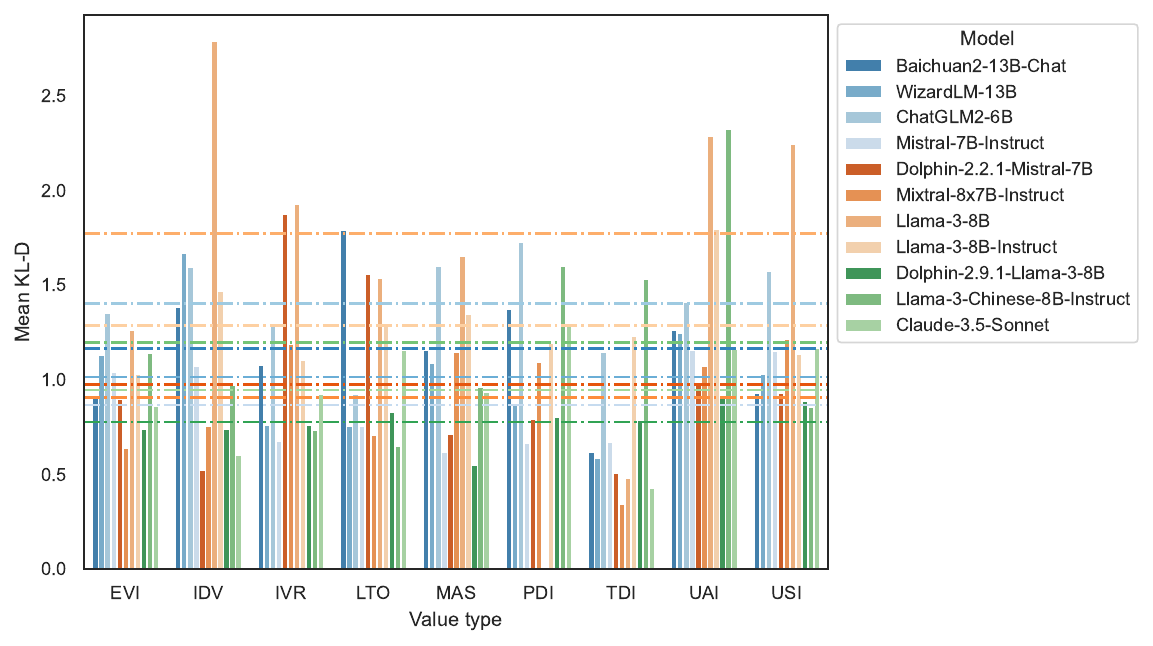}
    }
    \caption{Mean KL-Divergence of LLM candidates on US and CN 9-value dimensions. Horizontal line: average KL-D. Dolphin-2.9.1-Llama-3-8B and Mistral-7B-Instruct show consistent alignment across both cultures. \textit{Adapted from \citet{Liu25Towards}.}}
    \label{fig:val_mean}
\end{figure*}

\paragraph{Full-parameter fine-tuning versus reinforcement learning} Systematic comparison of alignment methodologies reveals technique-specific performance profiles.

In the context of Full-Parameter Fine-Tuning (FFT), models that have been fine-tuned using automatically generated diverse datasets, such as \textbf{Dolphin-2.9.1-Llama-3-8B}, demonstrate a markedly superior \textbf{average preference distribution accuracy} (Figure \ref{fig:val_mean}). Additionally, these models exhibit an enhanced capability to preserve \textbf{cross-cultural value variation} when compared to those aligned through Reinforcement Learning from Human Feedback (RLHF). This observation suggests that the diversity of the data, rather than the density of human preference annotations, is the primary factor influencing distributional fidelity.

Reinforcement Learning from Human Feedback (RLHF), specifically through the application of Supervised Fine-Tuning (SFT) combined with RLHF pipelines (such as Llama-3-8B-Instruct), results in significant reductions in systemic inconsistencies, as evidenced by metrics such as Frequency of Failure (FF) and Conflict in Value Expression (CV). Furthermore, this approach demonstrates superior internal coherence across evaluation contexts in both the United States and China. The findings suggest that iterative human feedback tends to optimize for consistency rather than distributional coverage.

This divergence suggests that optimal alignment strategies may require \textbf{hybrid approaches}: FFT for distributional coverage combined with RLHF for consistency refinement.

\subsubsection{Multilingual Training and Region-Specific Optimization: China Development Paradigm}
\paragraph{Impact of multilingual data integration} China-origin model development frequently incorporates substantial multilingual training data, providing a natural experiment in cross-cultural alignment:
\begin{itemize}
    \item Context-specific performance gains: Enhanced multilingual data percentages (as in \textbf{Llama-3-Chinese-8B-Instruct}) yield targeted improvements in Chinese cultural context performance and reduced insensitivity in Chinese evaluations
    \item \textbf{Tradeoff considerations}: Improvements remain context-specific, with limited transfer to U.S. cultural alignment, suggesting that multilingual data integration primarily benefits represented language-culture pairs
\end{itemize}

\paragraph{Base model foundations and alignment efficacy} Analysis of alignment intervention efficacy reveals the critical role of pretrained base models:
\begin{itemize}
    \item \textbf{Sequential alignment on base models}: Specialized techniques applied to base models yield substantially greater improvements than comparable interventions on instruction-tuned variants
    \item \textbf{Bias profiles across development stages}: Base models consistently exhibit reduced demographic bias compared to their instruction-tuned successors, suggesting that alignment processes (both SFT and RLHF) may inadvertently concentrate preference distributions while pursuing coherence
\end{itemize}

This finding has significant implications for \textbf{alignment strategy sequencing}: optimal workflows may involve comprehensive distributional alignment at the base model stage, followed by targeted consistency refinement through instruction tuning.

\subsubsection{Advanced Multi-Stage Reasoning Approaches}
\begin{table*}[]
\centering
\resizebox{\textwidth}{!}{%
\begin{tabular}{l|rrrr|rrrr|rrrr}
\toprule
\multicolumn{1}{c|}{\textbf{Model}} &
  \multicolumn{4}{c|}{\textbf{GLM-4-air}} &
  \multicolumn{4}{c|}{\textbf{GPT-4o}} &
  \multicolumn{4}{c}{\textbf{Doubao-1.5-pro}} \\ 
  \midrule \midrule
\textbf{Avg. Metrics} &
  \textbf{ACC} (\%) &
  \textbf{1-JSD} $\uparrow$ &
  \textbf{EMD} $\downarrow$ &
  \textbf{$\kappa$} $\uparrow$ &
  \textbf{ACC} (\%) &
  \textbf{1-JSD} $\uparrow$ &
  \textbf{EMD} $\downarrow$ &
  \textbf{$\kappa$} $\uparrow$ &
  \textbf{ACC} (\%) &
  \textbf{1-JSD} $\uparrow$ &
  \textbf{EMD} $\downarrow$ &
  \textbf{$\kappa$} $\uparrow$ \\
    \midrule
\textbf{Demo.+Ideo.} &
  25.49 &
  0.3539 &
  0.0741 &
  0.02 &
  32.30 &
  0.4075 &
  0.0755 &
  0.09 &
  30.75 &
  0.4563 &
  0.0685 &
  0.02 \\
\textbf{Demo.+Ideo.+Opinion} &
  25.06 &
  0.4313 (0.6) &
  0.0669 &
  0.11 &
  33.07 &
  0.4069 &
  0.0739 &
  0.09 &
  31.45 &
  0.4723 &
  0.0703 &
  0.08 \\
\textbf{\citet{zhao-etal-2024-worldvaluesbench}} &
  26.98 &
  0.3814 &
  0.0452 &
  0.05 &
  36.96 &
  0.4654 &
  0.0610 &
  0.12 &
  24.23 &
  0.3584 &
  0.0364 &
  0.05 \\
    \midrule 
\textbf{MARK (Ours)} &
  \textbf{33.69} &
  \textbf{0.4348} &
  0.0887 &
  \textbf{0.15} &
  \textbf{38.11} &
  \textbf{0.4879} &
  0.0826 &
  \textbf{0.15} &
  \textbf{46.98} &
  \textbf{0.5195} &
  0.0561 &
  \textbf{0.09}\\
  \bottomrule
\end{tabular}%
}

\caption{MARK simulation performance on U.S. social‐survey data, evaluated under both global‐distribution and sampled‐distribution settings. `\textit{Avg.}' denotes the overall mean metric values, (value) denotes the p-value of results between baseline and MARK larger than 0.05. MARK shows improvements on sampled distributions by achieving the \textbf{highest simulation accuracy} while \textbf{maintaining low divergences}, with most improvements being statistically significant. It also shows \textbf{generalization to global distributions} with unseen demographics. \textit{Adapted from \citet{liu2025demographics}.}}
\label{tbl:chara_res}
\end{table*}

Recent innovations demonstrate that multi-stage personality-driven cognitive reasoning frameworks can enhance cultural value survey simulation accuracy. MARK (Multi-stAge Reasoning frameworK), inspired by MBTI personality type dynamics, outperforms existing baselines by 10\% accuracy and reduces divergence between model predictions and human preferences. This suggests that incorporating psychological frameworks and multi-stage reasoning can improve cross-cultural alignment beyond traditional demographic-based approaches, representing a promising direction for both China and Western development paradigms.

\subsection{Overview: Convergent Challenges and Divergent Solutions in China-Western LLM Development}
The cross-cultural audit reveals a complex landscape of \textbf{convergent fundamental challenges} and \textbf{divergent methodological responses} across China and Western LLM development ecosystems.

\begin{figure}
    \centering
    \includegraphics[width=\linewidth]{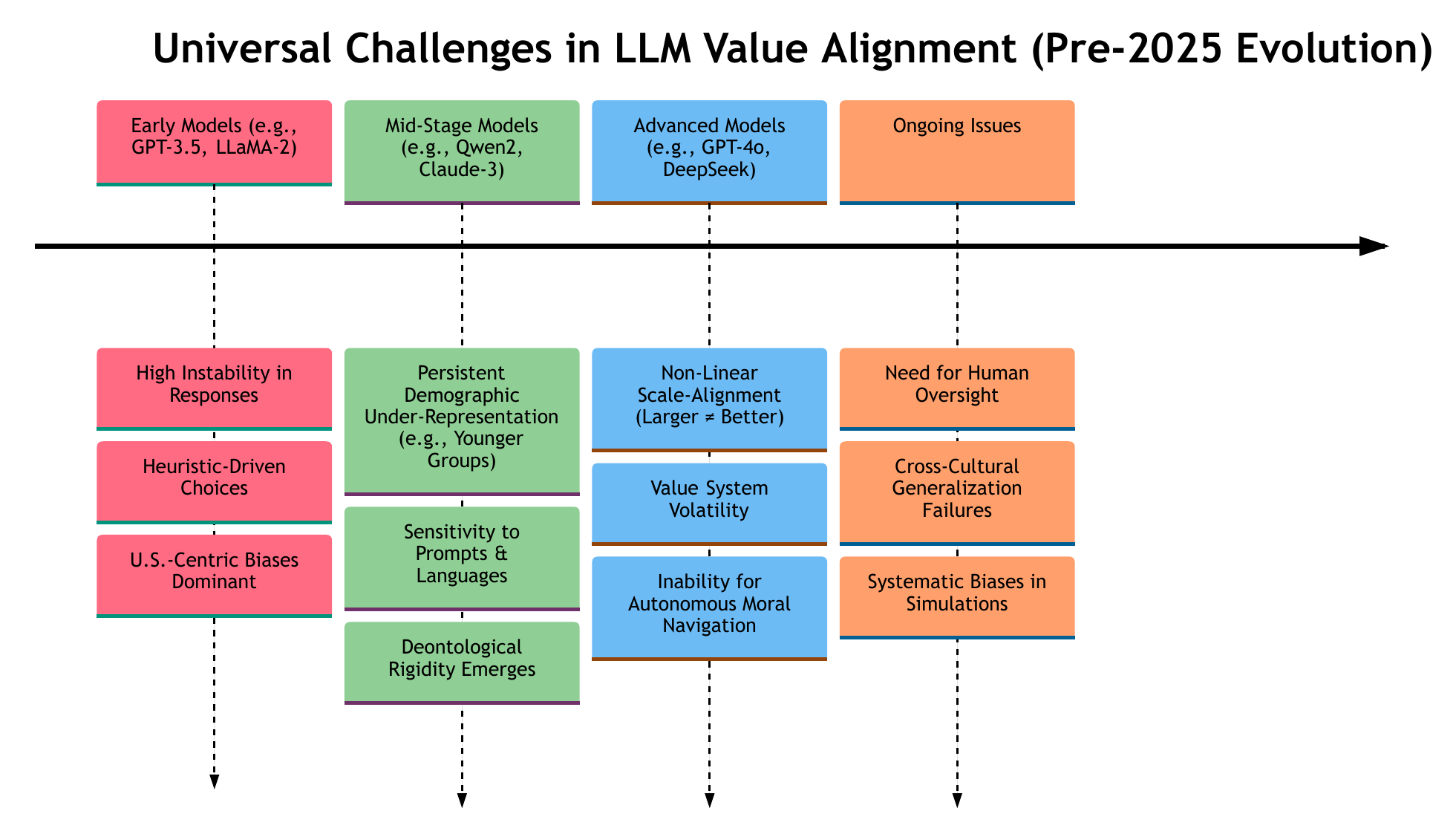}
    \caption{\textbf{Summary of Universal Challenges}: a timeline summarizing common challenges across China-West LLMs, evolving from early to advanced models.}
    \label{fig:uni_cha}
\end{figure}

\paragraph{Universal challenges} As summarized in Figure \ref{fig:uni_cha}, both regional contexts demonstrate: (1) fundamental instability in value systems despite surface-level consistency; (2) systematic under-representation of younger demographics; (3) non-linear relationships between model scale and cultural alignment quality; and (4) inherent tradeoffs between distributional fidelity and internal consistency.

\paragraph{Regional pathway divergence} China-origin development increasingly emphasizes multilingual data integration and region-specific optimization, yielding models with strong in-context performance but limited cross-cultural transfer. Western development demonstrates greater architectural experimentation (e.g., MoE configurations, Mistral innovations) but exhibits persistent U.S.-centric biases. Neither pathway has yet achieved robust, generalizable cross-cultural alignment.

\paragraph{Methodological implications} The findings suggest that next-generation alignment strategies must transcend current regional paradigms, integrating: (1) diverse data coverage (FFT strengths) with consistency mechanisms (RLHF strengths); (2) base model distributional alignment preceding instruction tuning; (3) explicit demographic representation objectives alongside cultural value targets; (4) multi-stage personality-driven reasoning approaches; and (5) systematic audit frameworks that evaluate both within-culture accuracy and cross-cultural preservation.

These results provide an empirical foundation for developing \textbf{harmonized international AI governance frameworks} that acknowledge regional innovation strengths while addressing shared fundamental challenges in responsible AI development.

\section{Technology for Good: Social Impact, LLM Governance, and Policy Translation}

\begin{figure}
    \centering
    \includegraphics[width=0.85\linewidth]{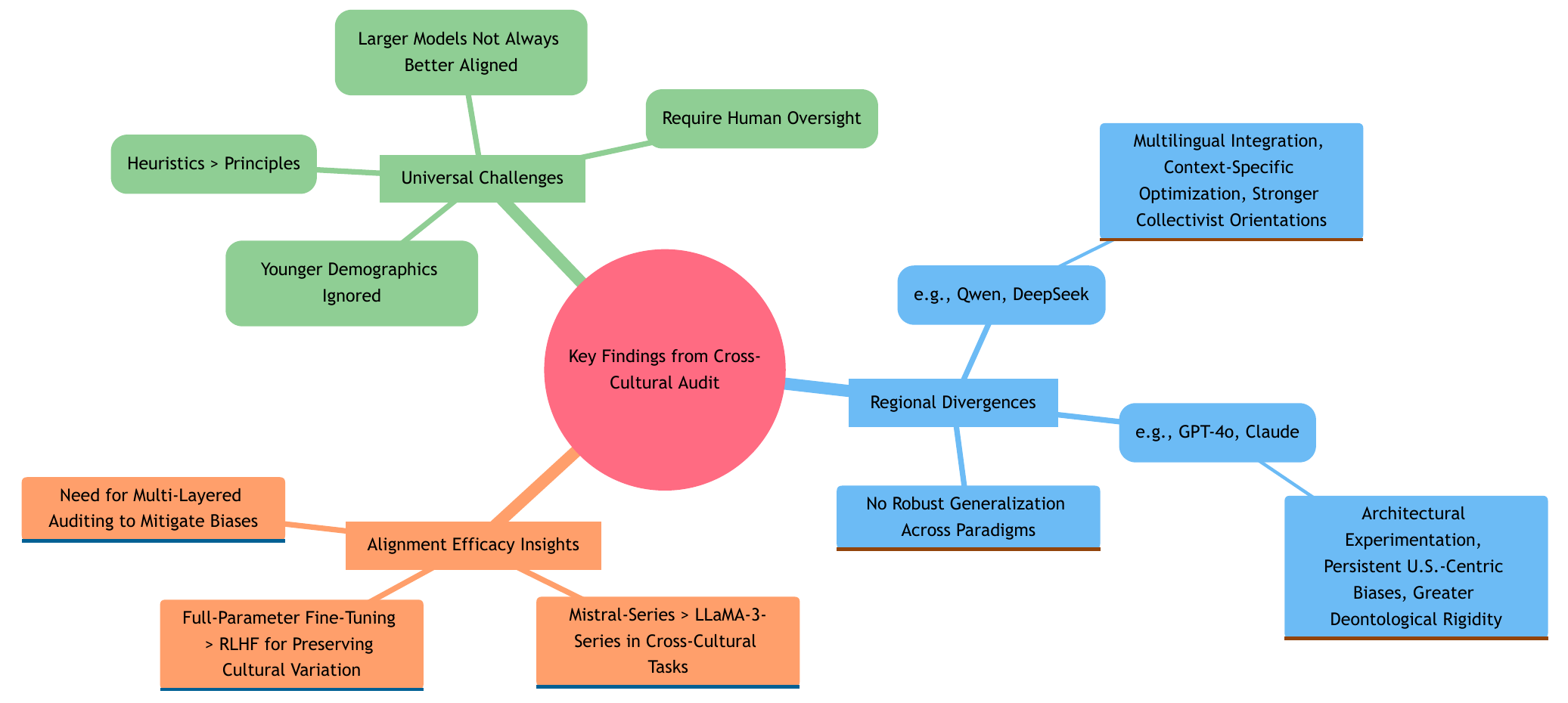}
    \caption{\textbf{Summary of Key Empirical Findings}: This mindmap captures the core findings from the cross-cultural audit, emphasizing universal challenges and divergences as foundations for governance discussions.}
    \label{fig:fi-all}
\end{figure}

This section articulates the strategic applications of the computational auditing platform, emphasizing governance protocols, policy implications, and institutional frameworks derived from the comparative analysis of China-Western LLM value alignment. As presented in Figure \ref{fig:fi-all}, we position these findings within broader debates on responsible AI governance, cross-cultural technology policy, and the democratization of computational social science methods.

\subsection{Computational Infrastructure for Responsible AI Governance: Public Policy and Academic Applications}
The Multi-Layered Auditing Platform represents a methodological contribution to the emerging field of algorithmic governance, enabling the transition from diagnostic evaluation to actionable policy interventions with measurable social impact.

\subsubsection{Case Study 1: Deliberative Democracy Through Computational Simulation—Policy Consultation at Scale}

\paragraph{Theoretical Framework and Policy Challenge}
Democratic governance increasingly confronts the challenge of incorporating diverse public perspectives into policy formation under time and resource constraints. Traditional deliberative mechanisms—public consultations, citizen assemblies, and survey research—face inherent limitations in scale, cost, and representational fidelity \citep{fishkin2011people, moore_democratic_2014}. Our integration of the \textbf{Diversity-Enhanced Framework (DEF)} with \textbf{First-Token Fine-Tuning} offers a novel computational approach to this democratic deficit.

\paragraph{Methodological Innovation and Governance Application}
This framework enables rapid, high-fidelity simulation of public attitudes and value trade-offs regarding proposed policies (e.g., energy transition measures, public health interventions, social welfare reforms) prior to legislative implementation. By achieving high \textbf{distributional fidelity}, these LLM-based simulations provide policymakers with \textbf{cost-effective, population-representative insights} into how demographically diverse populations—such as those in the United States versus China—might respond to policy proposals under consideration.

The framework advances beyond conventional survey prediction by specializing LLMs for distributional simulation rather than individual-level forecasting. Through a fine-tuning methodology leveraging first-token probability optimization, we minimize divergence between predicted and empirically observed country-level response distributions. This technical innovation yields substantial improvements in predictive accuracy over zero-shot baselines (e.g., 34.3\% increase for Llama3-8B-Instruct on aggregate measures), thereby enhancing the reliability of computational methods in social science research and evidence-based policymaking.

\paragraph{Policy Implications and Democratic Theory}
This application contributes to ongoing debates regarding the role of computational methods in deliberative democracy \citep{mercier_reasoning_2012}. While not replacing authentic public participation, these simulations can inform policy design phases, identify potential areas of public concern, and enable rapid iteration of policy proposals before resource-intensive implementation. This represents a form of "computational deliberation" that complements rather than substitutes traditional democratic processes.

\subsection{Case Study 2: Algorithmic Accountability Through Mechanistic Interpretability—Auditing AI Decision-Making Systems}
\paragraph{Governance Challenge and Accountability Framework}
As large language models increasingly inform high-stakes decisions across healthcare, criminal justice, and financial services, the opacity of their decision-making processes presents fundamental challenges to democratic accountability and procedural justice \citep{Citron2014TheSS, Selbst19Fairness}. The Multi-stAge Reasoning frameworK (MARK) addresses this accountability deficit by providing mechanistic, theory-grounded explanations for simulated value judgments.

\paragraph{Methodological Contribution to AI Transparency}
MARK operationalizes computational interpretability through cognitive modeling informed by personality psychology (Myers-Briggs Type Indicator framework) and stress response theory. This approach yields \textbf{personality-driven explanations} for value-laden choices, achieving superior accuracy compared to baseline simulation methods while demonstrating robust generalization to Chinese cultural contexts. Critically, the framework enables domain experts to \textbf{scrutinize and validate the reasoning trajectories} underlying LLM outputs, thereby operationalizing the principle of "algorithmic due process."

\paragraph{Institutional and Regulatory Implications}
This work contributes to the emerging regulatory landscape surrounding AI explainability requirements, such as the EU AI Act \citep{act2024eu} and the proposed U.S. Algorithmic Accountability Act. By providing granular, theory-informed explanations, MARK offers a potential compliance pathway for organizations subject to explainability mandates while advancing scholarly understanding of how cultural context shapes algorithmic reasoning. The cross-cultural validation is particularly significant given the comparative paucity of AI governance research outside Western contexts.

\subsection{Case Study 3: Normative Stability in Sequential Decision-Making—A Safety Protocol for Value Alignment}
\paragraph{Risk Assessment Framework}
The deployment of LLMs in autonomous decision-making systems raises fundamental questions about moral consistency and normative reliability over extended interaction sequences. The \textbf{Ethical Dilemma Corpus} provides an empirical foundation for assessing this previously under-examined risk vector in AI safety research.

\paragraph{Key Empirical Findings and Theoretical Implications}
Our sequential stability analysis reveals that LLMs exhibit \textbf{non-transitive and temporally shifting moral preferences} across multi-stage ethical scenarios. Rather than demonstrating stable normative principles analogous to human moral reasoning \citep{Kohlberg81-KOHTPO-6, haidt2013righteous}, models rely on \textbf{context-dependent heuristics and statistical pattern matching}. This instability manifests across both Chinese (DeepSeek) and Western (Llama-series) models, suggesting a fundamental architectural limitation rather than a culture-specific training artifact.

\paragraph{Regulatory Guidance and Safety Protocols}
These findings yield critical safety guidance for AI governance: \textbf{LLMs should not be authorized for autonomous, sequential ethical decision-making} in high-stakes domains absent robust human oversight mechanisms. This evidence-based constraint informs emerging regulatory frameworks, including sector-specific guidance for healthcare AI, autonomous vehicles, and automated content moderation systems. The findings support a "human-in-the-loop" governance model rather than fully autonomous algorithmic decision-making for value-laden judgments.

\subsection{Strategic Framework for LLM Governance}
\begin{figure}
    \centering
    \includegraphics[width=0.6\linewidth]{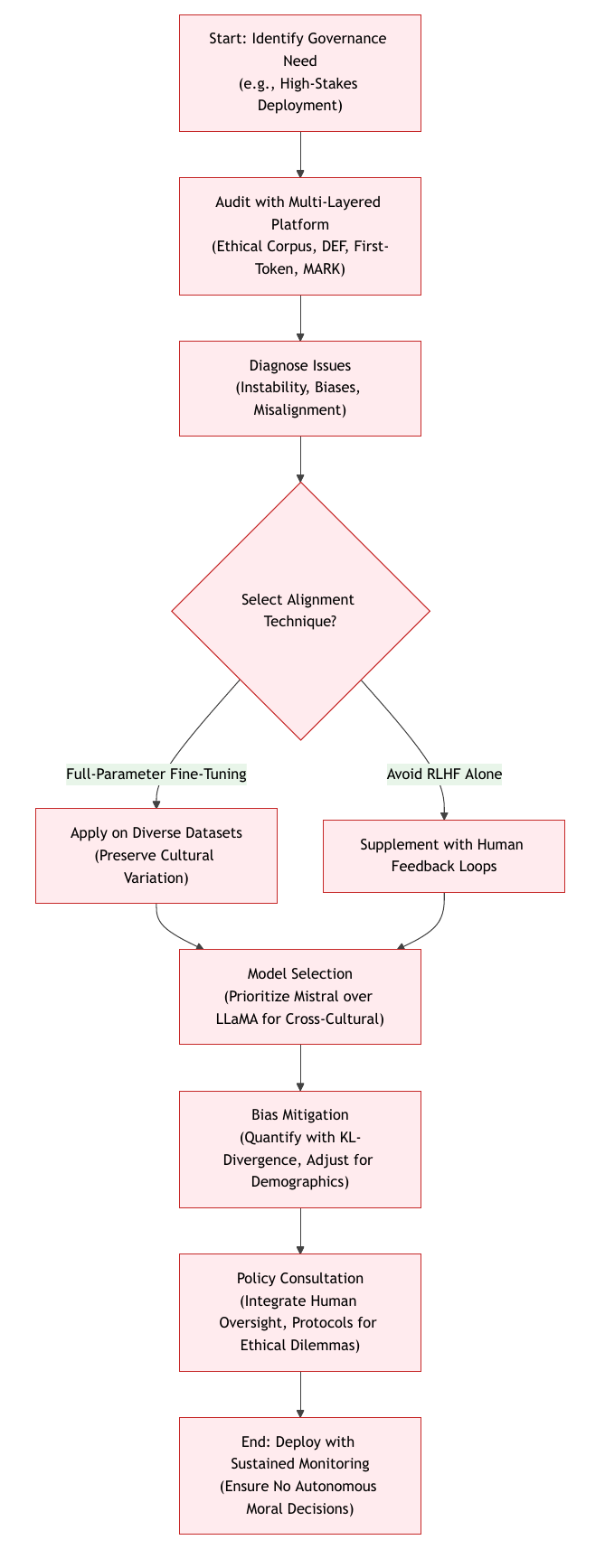}
    \caption{\textbf{Actionable Protocols for AI Governance Suggestions}: This flowchart outlines practical suggestions for model selection, bias mitigation, and policy consultation, as derived from the paper's emphasis on evidence-based governance.}
    \label{fig:goven-pip}
\end{figure}

The comparative audit generates actionable strategic recommendations for institutional LLM governance, informing model procurement decisions, bias mitigation protocols, and accountability mechanisms—particularly relevant for organizations operating across diverse cultural contexts (illustrated in Figure \ref{fig:goven-pip}).

\subsubsection{Strategic Model Selection: Empirically-Grounded Procurement Decisions}
\paragraph{Evidence-Based Architecture Assessment}
Institutional governance of AI systems requires moving beyond vendor claims to empirically validated performance metrics. Our cross-cultural value alignment audit establishes that Mistral-series models demonstrate statistically significant superior performance relative to Llama-3-series architectures across both U.S. and Chinese cultural contexts (p < 0.01). This finding provides quantitative evidence to inform risk-adjusted model selection for organizations deploying LLMs in culturally diverse markets or multilingual contexts.

\paragraph{Implications for Technology Procurement Policy}
This comparative finding advances procurement policy by establishing culture-agnostic performance benchmarks. Organizations can leverage these empirical comparisons to justify model selection decisions, establish baseline performance requirements in vendor contracts, and implement evidence-based standards for model evaluation. This represents a shift from opaque, proprietary benchmarks to transparent, replicable evaluation methodologies.

\subsubsection{Governance Architecture: Transparency-Enabling Alignment Techniques}
\paragraph{Comparative Evaluation of Alignment Methods}
Effective AI governance requires not only aligned outputs but also interpretable alignment processes that enable institutional oversight. Our research demonstrates that \textbf{explicit value constraint specification} substantially outperforms implicit few-shot learning for guiding normative judgments. Explicit prompting strategies yield superior transparency because they articulate constraints directly, whereas few-shot examples require inferential leaps that obscure the operative value framework.

\paragraph{Performance Benchmarks and Best Practices}
Leading models employing explicit preference alignment achieve substantial performance gains, with Claude-3.5-Sonnet reaching 83.1\% accuracy and Llama-3-70B achieving 77.8\%—both substantially exceeding their baseline zero-shot performance. These findings support governance protocols that mandate \textbf{explicit value specification} in high-stakes applications, enabling clearer accountability chains and more straightforward auditing processes.

\subsubsection{Accountability Mechanisms: Quantifying and Rectifying Demographic Bias}
\paragraph{Empirical Bias Characterization}
Algorithmic accountability requires moving beyond abstract fairness principles to concrete, measurable bias profiles. Leveraging the \textbf{Diversity-Enhanced Framework's demographic preference mapping}, we identify systematic cross-cultural patterns in representational bias:
\begin{enumerate}
    \item \textbf{U.S. context}: Optimal alignment configurations systematically favor male demographic profiles aged 30-49 or over 50, indicating potential underrepresentation of female and younger perspectives
    \item \textbf{Chinese context}: Superior alignment emerges for female personas aged 30-49, suggesting inverse gender skew relative to Western models
    \item \textbf{Cross-cultural pattern}: Both cultural contexts demonstrate inadequate representation of preferences associated with individuals under age 29, indicating a systematic generational bias
\end{enumerate}

\paragraph{Targeted Mitigation Strategies}
These empirically grounded bias profiles inform differentiated mitigation strategies. Rather than one-size-fits-all debiasing approaches, organizations should implement \textbf{culturally-calibrated interventions} addressing the specific demographic misalignments identified in their operational contexts. This represents a shift from generic fairness interventions to precision bias mitigation informed by rigorous empirical assessment.

\begin{figure}
    \centering
    \includegraphics[width=0.8\linewidth]{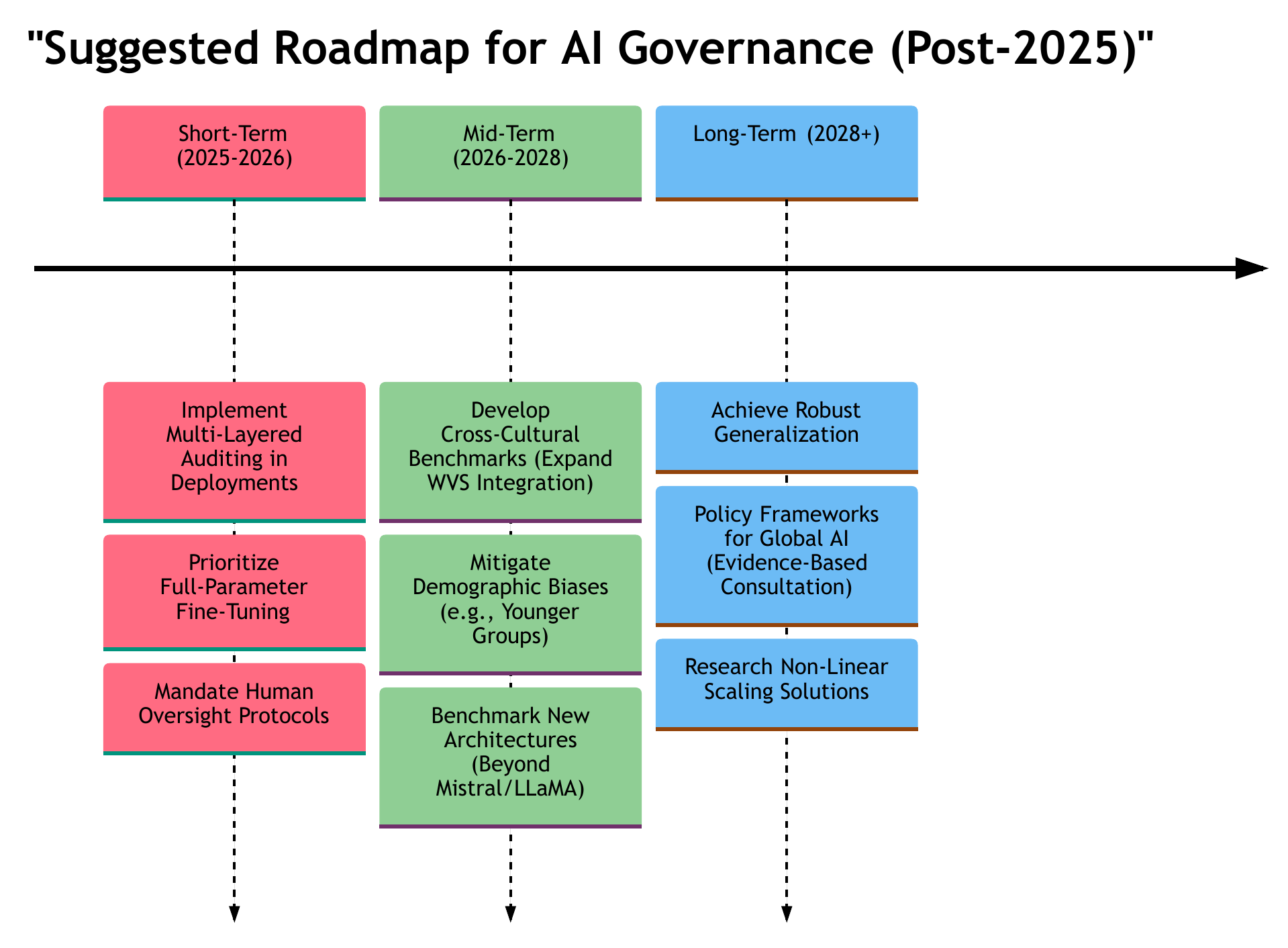}
    \caption{\textbf{Future Directions and Suggestions for Responsible AI}: This timeline projects suggestions for advancing AI governance based on the findings, focusing on short-term actions and long-term research needs.}
    \label{fig:road}
\end{figure}

\subsubsection{Technical Compliance Pathways: Engineering Solutions for Measurable Standards}
Our evaluation of alignment techniques yields specific technical recommendations for organizations seeking to operationalize cultural sensitivity and fairness standards (timelined in Figure \ref{fig:road}):
\begin{itemize}
    \item \textbf{First-Token Alignment}: This specialized fine-tuning approach demonstrates substantial accuracy improvements and validates the technical feasibility of correcting cross-cultural distributional misalignments through targeted architectural interventions
    \item \textbf{Full-Parameter Fine-Tuning (FFT) vs. RLHF}: Systematic comparison reveals that FFT using automatically enhanced training data achieves superior performance in preserving cultural variation and improving average preference distributions relative to Reinforcement Learning from Human Feedback on human-annotated data
    \item \textbf{Multilingual Data Integration}: Increased multilingual training data—a characteristic feature of Chinese model development approaches—demonstrably improves performance in culture-specific alignment tasks and reduces cultural insensitivity, offering potential lessons for Western model development
\end{itemize}

These technical pathways enable organizations to translate abstract fairness commitments into measurable engineering interventions. By establishing quantitative benchmarks for cultural alignment and demonstrating effective mitigation techniques, this research provides actionable compliance pathways for organizations navigating emerging AI governance regulations.

\section{Discussion}
This study deployed a \textbf{Multi-Layered Auditing Platform for Responsible AI} that synthesizes computational methodologies to systematically evaluate Large Language Model (LLM) behavior across cultural and temporal dimensions. The platform integrates four complementary analytical instruments: the \textbf{Ethical Dilemma Corpus}, which interrogates temporal stability and detects reliance on unstable heuristics through progressively intensifying ethical conflicts; the \textbf{Diversity-Enhanced Framework (DEF)}, which quantifies representational fidelity by benchmarking model outputs against empirical human survey distributions in U.S. and Chinese contexts; \textbf{First-Token Probability Alignment}, which enhances distributional simulation accuracy through refined calibration against ground-truth human preferences; and the \textbf{Multi-stAge Reasoning frameworK (MARK)}, which elucidates mechanistic, personality-driven explanations to strengthen accountability and interpretability. This comprehensive architecture enables dynamic, multi-dimensional scrutiny of China-Western LLM ecosystems, addressing critical gaps in cross-cultural AI governance research.

\subsection{Empirical Findings from Cross-Cultural Auditing}
Answering \textit{\textbf{RQ1}: Temporal Stability, Value Rigidity, and Sequential Moral Reasoning}, the comparative analysis highlights that both Chinese and Western LLMs demonstrate pronounced \textbf{fixed value hierarchies}, consistently prioritizing Truth over Loyalty and Community over Individual welfare. Larger models tend to adopt deontological frameworks that maintain ethical commitments despite adverse outcomes. However, the Ethical Dilemma Corpus reveals that LLMs rely on \textbf{heuristic-driven, context-dependent statistical imitation} rather than principled reasoning, limiting their ability for autonomous ethical decision-making. While advanced models like Claude-3.5-Sonnet and Qwen2-72B show improved task comprehension, they exhibit varied stability profiles across dilemma steps. The moral foundation of \textbf{Care} remains consistently stable, while \textbf{Authority} prioritization becomes less stable. Importantly, moral volatility is not confined to any region, with both DeepSeek and Llama showing instability, whereas GLM-4-Air and Claude demonstrate stable value maintenance, suggesting that architectural choices are more crucial than geographic origin in ethical consistency.

For \textit{\textbf{RQ2}: Cultural Fidelity, Systematic Bias, and Architectural Performance Differentials}, a quantitative assessment of LLM-human preference alignment reveals significant performance disparities between U.S. and Chinese cultural contexts, with \textbf{Mistral-series} models outperforming the others in value alignment. Preference bias analysis uncovers demographic misalignment: U.S. context favors \textbf{males aged 30-49 or 50+}, while the Chinese context skews towards \textbf{females aged 30-49}. Notably, both cultures show a critical under-representation of preferences from individuals \textbf{under 29}, highlighting a generational bias in current LLMs.

To resolve \textit{\textbf{RQ3}: Developmental Trajectories, Alignment Innovation, and Cross-Regional Strategy Divergence}, a longitudinal analysis of LLM development shows that model evolution and cultural alignment efficacy have complex, non-linear relationships. Although generational advancements reduce insensitivity failures, \textbf{increasing model scale does not guarantee better alignment quality or cultural representation}, challenging assumptions about parameter count as a key ethical performance factor. Specialized First-Token Alignment methods provide significant accuracy improvements over zero-shot techniques. Additionally, \textbf{Full-Parameter Fine-Tuning (FFT) on enhanced datasets generally outperforms Reinforcement Learning from Human Feedback (RLHF) trained on human-analyzed data} in maintaining preference distributions and cultural variation. A notable divergence exists between China and the West, with China's approach of \textbf{increasing multilingual training data} (e.g., Llama-3-Chinese-8B-Instruct) enhancing culture-specific alignment and reducing insensitivity in Chinese contexts, impacting global AI strategies balancing linguistic diversity and cultural fidelity.

\subsection{Strategic Implications for Responsible AI Governance}
\paragraph{Policy Validation and Transparency Mechanisms}
The empirical platform developed through DEF and First-Token Fine-Tuning provides \textbf{cost-effective} tools for regulatory bodies to assess population responses to policies across various cultures (U.S. vs. China). This approach enhances the speed of social science research and supports evidence-based governance. The MARK Framework offers \textbf{mechanistic, personality-driven insights} into value choices, allowing experts to validate AI reasoning and ensuring necessary human oversight in AI-driven decisions. Given that current LLMs rely on unstable heuristics rather than consistent ethical principles, a foundational safety protocol is essential. \textbf{LLMs should not be allowed to make autonomous ethical decisions in high-stakes situations} and require ongoing human supervision, particularly in sectors like healthcare, criminal justice, and public policy where decisions have serious social implications.

\paragraph{Model Selection and Bias Mitigation}
Evidence-based governance requires prioritizing empirical alignment data, particularly highlighting the \textbf{superior performance of Mistral-series architectures over Llama-3} in cross-cultural contexts. Accountability frameworks should mandate explicit alignment techniques that enforce \textbf{explicit ethical policies or value constraints} rather than relying on implicit learning. Targeted bias mitigation must address demographic skews, such as \textbf{U.S. male/older-age bias versus Chinese female/middle-aged bias}, while correcting \textbf{under-representation of the under-29 age group}. Specialized fine-tuning methodologies like First-Token Alignment offer efficient solutions for achieving fairness and cultural sensitivity, indicating a need to shift resources from RLHF to FFT in alignment research.

\section{Conclusion}
This study demonstrates that responsible AI development in an interconnected world requires moving beyond single-culture optimization toward genuine cross-cultural competence. The proposed computational auditing platform: \textbf{Multi-
Layered Auditing Platform for Responsible AI}, provides methodological foundations for this transition, enabling systematic evaluation, transparent accountability, and evidence-based governance across diverse cultural contexts.

As LLMs shape global communication and public discourse, aligning them with diverse human values is essential for equitable technological development. Our findings establish that this alignment cannot be assumed from model scale, assumed from training data volume, or inherited from single-culture optimization. It must be systematically engineered, rigorously audited, and continuously refined through transparent, reproducible methodologies that acknowledge both universal principles and cultural particularity.

The comparative China-Western analysis reveals that technological innovation serving the global good requires integrating diverse epistemological approaches, respecting cultural specificity while pursuing shared ethical standards, and maintaining a humble recognition of current limitations alongside an ambitious pursuit of improvement. By establishing empirical foundations for evidence-based cross-cultural AI governance, this research contributes to building artificial intelligence systems that genuinely serve humanity in its full diversity—systems that enhance rather than erase cultural richness, that amplify rather than silence marginalized voices, and that operate transparently under sustained human guidance rather than claiming autonomous moral authority.

The path toward culturally competent, socially beneficial AI systems remains long and technically demanding. However, through systematic evaluation frameworks, transparent accountability mechanisms, and international collaborative governance, we can work toward artificial intelligence that serves not merely dominant cultural paradigms but the full spectrum of human values, experiences, and aspirations across our diverse global community. This study provides both the methodological tools and empirical insights to guide that essential journey.

\section*{Limitations}
\paragraph{Conceptual Framework Limitations} 
The foundational moral frameworks employed (Kidder's Ethical Dilemmas, Moral Foundations Theory, Schwartz's Theory of Basic Values) inherently privilege Western-centric moral constructs, potentially underrepresenting collectivist ethics such as Confucian \textit{ren} or Ubuntu \textit{ubuntu} that are crucial in non-Western contexts. Future research must systematically integrate culture-specific ethical dimensions through sustained collaboration with regional cultural experts and philosophers representing diverse philosophical traditions.

\paragraph{Methodological Design Constraints}
LLMs exhibit pronounced sensitivity to prompt presentation, and the Ethical Dilemma Corpus framework's linearly intensifying scenarios cannot model non-linear escalations (e.g., de-escalation through negotiation), while MARK's reliance on LLM-generated personality predictions and the empirical limitations of MBTI constrain generalizability. Future methodologies should incorporate validated psychological instruments, explore diverse scenario progression patterns, and develop comprehensive prompt variation testing protocols.

\paragraph{Empirical Scope Limitations}
The focus on U.S. and Chinese cultural contexts using English and Chinese prompts does not capture broader global cultural diversity, while First-Token Alignment models remain highly specialized for survey response prediction rather than general-purpose applications. Expanding research to encompass diverse cultural contexts, languages, and real-world deployment scenarios is essential for establishing cross-cultural generalizability.


\paragraph{Governance Transparency Barriers}
The lack of transparency regarding training data sources in commercial and open LLMs (e.g., Mistral, Llama-3) represents a significant constraint on auditability, particularly in assessing the impact of multilingual data integration—a key feature of Chinese AI innovation—on model behavior. Future governance frameworks must mandate comprehensive training data disclosure and establish standardized transparency protocols for cross-cultural AI auditing.

\section*{Acknowledgments}
We thank the anonymous reviewers for their valuable feedback and constructive comments that helped improve this work. We are grateful to Yuchu Tian and Caicai Guo for their insightful discussions and suggestions.

This work was supported by the National Key Research and Development Program of China under Grants 2022YFC3300801, the Research Project of China Publishing Promotion Association under Grants 2025ZBCH-JYYB17, and the Natural Science Foundation of Hubei Province (CN) under Grant No. 2025AFB078.

\bibliography{custom}

\begin{thebibliography}{45}
\providecommand{\natexlab}[1]{#1}
\providecommand{\url}[1]{\texttt{#1}}
\expandafter\ifx\csname urlstyle\endcsname\relax
  \providecommand{\doi}[1]{doi: #1}\else
  \providecommand{\doi}{doi: \begingroup \urlstyle{rm}\Url}\fi

\bibitem[Act(2024)]{act2024eu}
EU~Artificial~Intelligence Act.
\newblock The eu artificial intelligence act.
\newblock \emph{European Union}, 2024.

\bibitem[Agarwal et~al.(2024{\natexlab{a}})Agarwal, Tanmay, Khandelwal, and Choudhury]{Agarwal24Ethical}
Utkarsh Agarwal, Kumar Tanmay, Aditi Khandelwal, and Monojit Choudhury.
\newblock Ethical reasoning and moral value alignment of llms depend on the language we prompt them in.
\newblock In Nicoletta Calzolari, Min{-}Yen Kan, V{\'{e}}ronique Hoste, Alessandro Lenci, Sakriani Sakti, and Nianwen Xue (eds.), \emph{Proceedings of the 2024 Joint International Conference on Computational Linguistics, Language Resources and Evaluation, {LREC/COLING} 2024, 20-25 May, 2024, Torino, Italy}, pp.\  6330--6340. {ELRA} and {ICCL}, 2024{\natexlab{a}}.
\newblock URL \url{https://aclanthology.org/2024.lrec-main.560}.

\bibitem[Agarwal et~al.(2024{\natexlab{b}})Agarwal, Tanmay, Khandelwal, and Choudhury]{agarwal-etal-2024-ethical}
Utkarsh Agarwal, Kumar Tanmay, Aditi Khandelwal, and Monojit Choudhury.
\newblock Ethical reasoning and moral value alignment of {LLM}s depend on the language we prompt them in.
\newblock In \emph{Proceedings of the 2024 Joint International Conference on Computational Linguistics, Language Resources and Evaluation (LREC-COLING 2024)}, pp.\  6330--6340, Torino, Italia, May 2024{\natexlab{b}}. ELRA and ICCL.
\newblock URL \url{https://aclanthology.org/2024.lrec-main.560/}.

\bibitem[Argyle et~al.(2022)Argyle, Busby, Fulda, Gubler, Rytting, and Wingate]{Lisa22Out}
Lisa~P. Argyle, Ethan~C. Busby, Nancy Fulda, Joshua Gubler, Christopher~Michael Rytting, and David Wingate.
\newblock Out of one, many: Using language models to simulate human samples.
\newblock \emph{CoRR}, abs/2209.06899, 2022.
\newblock \doi{10.48550/ARXIV.2209.06899}.
\newblock URL \url{https://doi.org/10.48550/arXiv.2209.06899}.

\bibitem[Argyle et~al.(2023)Argyle, Busby, Fulda, Gubler, Rytting, and Wingate]{Argyle_Busby_Fulda_Gubler_Rytting_Wingate_2023}
Lisa~P. Argyle, Ethan~C. Busby, Nancy Fulda, Joshua~R. Gubler, Christopher Rytting, and David Wingate.
\newblock Out of one, many: Using language models to simulate human samples.
\newblock \emph{Political Analysis}, 31\penalty0 (3):\penalty0 337–351, 2023.
\newblock \doi{10.1017/pan.2023.2}.

\bibitem[Bail(2024)]{Bail21Can}
Christopher~A. Bail.
\newblock Can generative ai improve social science?
\newblock \emph{Proceedings of the National Academy of Sciences}, 121\penalty0 (21):\penalty0 e2314021121, 2024.
\newblock \doi{10.1073/pnas.2314021121}.
\newblock URL \url{https://www.pnas.org/doi/abs/10.1073/pnas.2314021121}.

\bibitem[Cao et~al.(2023)Cao, Zhou, Lee, Cabello, Chen, and Hershcovich]{cao-etal-2023-assessing}
Yong Cao, Li~Zhou, Seolhwa Lee, Laura Cabello, Min Chen, and Daniel Hershcovich.
\newblock Assessing cross-cultural alignment between {C}hat{GPT} and human societies: An empirical study.
\newblock In \emph{Proceedings of the First Workshop on Cross-Cultural Considerations in NLP (C3NLP)}, pp.\  53--67, Dubrovnik, Croatia, May 2023. Association for Computational Linguistics.
\newblock \doi{10.18653/v1/2023.c3nlp-1.7}.
\newblock URL \url{https://aclanthology.org/2023.c3nlp-1.7/}.

\bibitem[Cao et~al.(2025)Cao, Liu, Arora, Augenstein, R{\"o}ttger, and Hershcovich]{cao-etal-2025-specializing}
Yong Cao, Haijiang Liu, Arnav Arora, Isabelle Augenstein, Paul R{\"o}ttger, and Daniel Hershcovich.
\newblock Specializing large language models to simulate survey response distributions for global populations.
\newblock In \emph{Proceedings of the 2025 Conference of the Nations of the Americas Chapter of the Association for Computational Linguistics: Human Language Technologies (Volume 1: Long Papers)}, pp.\  3141--3154, Albuquerque, New Mexico, April 2025. Association for Computational Linguistics.
\newblock ISBN 979-8-89176-189-6.
\newblock \doi{10.18653/v1/2025.naacl-long.162}.
\newblock URL \url{https://aclanthology.org/2025.naacl-long.162/}.

\bibitem[Christiano et~al.(2023)Christiano, Leike, Brown, Martic, Legg, and Amodei]{christiano2023deepreinforcementlearninghuman}
Paul Christiano, Jan Leike, Tom~B. Brown, Miljan Martic, Shane Legg, and Dario Amodei.
\newblock Deep reinforcement learning from human preferences, 2023.
\newblock URL \url{https://arxiv.org/abs/1706.03741}.

\bibitem[Citron \& Pasquale(2014)Citron and Pasquale]{Citron2014TheSS}
Danielle~Keats Citron and Frank~A. Pasquale.
\newblock The scored society: Due process for automated predictions.
\newblock \emph{Washington Law Review}, 89:\penalty0 1, 2014.
\newblock URL \url{https://api.semanticscholar.org/CorpusID:61799887}.

\bibitem[Dev et~al.(2022)Dev, Sheng, Zhao, Amstutz, Sun, Hou, Sanseverino, Kim, Nishi, Peng, and Chang]{dev-etal-2022-measures}
Sunipa Dev, Emily Sheng, Jieyu Zhao, Aubrie Amstutz, Jiao Sun, Yu~Hou, Mattie Sanseverino, Jiin Kim, Akihiro Nishi, Nanyun Peng, and Kai-Wei Chang.
\newblock On measures of biases and harms in {NLP}.
\newblock In \emph{Findings of the Association for Computational Linguistics: AACL-IJCNLP 2022}, pp.\  246--267, Online only, November 2022. Association for Computational Linguistics.
\newblock \doi{10.18653/v1/2022.findings-aacl.24}.
\newblock URL \url{https://aclanthology.org/2022.findings-aacl.24/}.

\bibitem[Dubey et~al.(2025)Dubey, Dailisan, and Mahajan]{dubey2025addressingmoraluncertaintyusing}
Rohit~K. Dubey, Damian Dailisan, and Sachit Mahajan.
\newblock Addressing moral uncertainty using large language models for ethical decision-making, 2025.
\newblock URL \url{https://arxiv.org/abs/2503.05724}.

\bibitem[Durmus et~al.(2024)Durmus, Nguyen, Liao, Schiefer, Askell, Bakhtin, Chen, Hatfield-Dodds, Hernandez, Joseph, Lovitt, McCandlish, Sikder, Tamkin, Thamkul, Kaplan, Clark, and Ganguli]{durmus2024measuringrepresentationsubjectiveglobal}
Esin Durmus, Karina Nguyen, Thomas~I. Liao, Nicholas Schiefer, Amanda Askell, Anton Bakhtin, Carol Chen, Zac Hatfield-Dodds, Danny Hernandez, Nicholas Joseph, Liane Lovitt, Sam McCandlish, Orowa Sikder, Alex Tamkin, Janel Thamkul, Jared Kaplan, Jack Clark, and Deep Ganguli.
\newblock Towards measuring the representation of subjective global opinions in language models, 2024.
\newblock URL \url{https://arxiv.org/abs/2306.16388}.

\bibitem[Emelin et~al.(2021)Emelin, Le~Bras, Hwang, Forbes, and Choi]{emelin-etal-2021-moral}
Denis Emelin, Ronan Le~Bras, Jena~D. Hwang, Maxwell Forbes, and Yejin Choi.
\newblock Moral stories: Situated reasoning about norms, intents, actions, and their consequences.
\newblock In \emph{Proceedings of the 2021 Conference on Empirical Methods in Natural Language Processing}, pp.\  698--718, Online and Punta Cana, Dominican Republic, November 2021. Association for Computational Linguistics.
\newblock \doi{10.18653/v1/2021.emnlp-main.54}.
\newblock URL \url{https://aclanthology.org/2021.emnlp-main.54/}.

\bibitem[Fishkin(2011)]{fishkin2011people}
J.S. Fishkin.
\newblock \emph{When the People Speak: Deliberative Democracy and Public Consultation}.
\newblock Oxford University Press, 2011.
\newblock ISBN 9780199604432.
\newblock URL \url{https://books.google.com/books?id=iNsUDAAAQBAJ}.

\bibitem[Forbes et~al.(2020)Forbes, Hwang, Shwartz, Sap, and Choi]{forbes-etal-2020-social}
Maxwell Forbes, Jena~D. Hwang, Vered Shwartz, Maarten Sap, and Yejin Choi.
\newblock Social chemistry 101: Learning to reason about social and moral norms.
\newblock In Bonnie Webber, Trevor Cohn, Yulan He, and Yang Liu (eds.), \emph{Proceedings of the 2020 Conference on Empirical Methods in Natural Language Processing (EMNLP)}, pp.\  653--670, Online, November 2020. Association for Computational Linguistics.
\newblock \doi{10.18653/v1/2020.emnlp-main.48}.
\newblock URL \url{https://aclanthology.org/2020.emnlp-main.48/}.

\bibitem[Hadar-Shoval et~al.(2024)Hadar-Shoval, Asraf, Mizrachi, Haber, and Elyoseph]{Shoval24Assessing}
Dorit Hadar-Shoval, Kfir Asraf, Yonathan Mizrachi, Yuval Haber, and Zohar Elyoseph.
\newblock Assessing the alignment of large language models with human values for mental health integration: Cross-sectional study using schwartz’s theory of basic values.
\newblock \emph{JMIR Mental Health}, 11, 2024.
\newblock ISSN 2368-7959.
\newblock \doi{https://doi.org/10.2196/55988}.
\newblock URL \url{https://www.sciencedirect.com/science/article/pii/S2368795924000350}.

\bibitem[Hagendorff(2020)]{Hagendorff20The}
Thilo Hagendorff.
\newblock The ethics of {AI} ethics: An evaluation of guidelines.
\newblock \emph{Minds Mach.}, 30\penalty0 (1):\penalty0 99--120, 2020.
\newblock \doi{10.1007/S11023-020-09517-8}.
\newblock URL \url{https://doi.org/10.1007/s11023-020-09517-8}.

\bibitem[Haidt(2013)]{haidt2013righteous}
J.~Haidt.
\newblock \emph{The Righteous Mind: Why Good People Are Divided by Politics and Religion}.
\newblock Knopf Doubleday Publishing Group, 2013.
\newblock ISBN 9780307455772.
\newblock URL \url{https://books.google.com/books?id=U21BxGfm3RUC}.

\bibitem[Hendrycks et~al.(2021)Hendrycks, Burns, Basart, Critch, Li, Song, and Steinhardt]{hendrycks2021ethics}
Dan Hendrycks, Collin Burns, Steven Basart, Andrew Critch, Jerry Li, Dawn Song, and Jacob Steinhardt.
\newblock Aligning ai with shared human values.
\newblock \emph{Proceedings of the International Conference on Learning Representations (ICLR)}, 2021.

\bibitem[Jiang et~al.(2022)Jiang, Hwang, Bhagavatula, Bras, Liang, Dodge, Sakaguchi, Forbes, Borchardt, Gabriel, Tsvetkov, Etzioni, Sap, Rini, and Choi]{jiang2022machineslearnmoralitydelphi}
Liwei Jiang, Jena~D. Hwang, Chandra Bhagavatula, Ronan~Le Bras, Jenny Liang, Jesse Dodge, Keisuke Sakaguchi, Maxwell Forbes, Jon Borchardt, Saadia Gabriel, Yulia Tsvetkov, Oren Etzioni, Maarten Sap, Regina Rini, and Yejin Choi.
\newblock Can machines learn morality? the delphi experiment, 2022.
\newblock URL \url{https://arxiv.org/abs/2110.07574}.

\bibitem[Jobin et~al.()Jobin, Ienca, and Vayena]{jobin19global}
Anna Jobin, Marcello Ienca, and Effy Vayena.
\newblock The global landscape of {AI} ethics guidelines.
\newblock 1\penalty0 (9):\penalty0 389--399.
\newblock ISSN 2522-5839.
\newblock \doi{10.1038/s42256-019-0088-2}.
\newblock URL \url{https://doi.org/10.1038/s42256-019-0088-2}.

\bibitem[Kidder(1996)]{kidder1996good}
R.M. Kidder.
\newblock \emph{How Good People Make Tough Choices: Resolving the Dilemmas of Ethical Living}.
\newblock A Fireside book. Simon \& Schuster, 1996.
\newblock ISBN 9780684818382.
\newblock URL \url{https://books.google.com/books?id=1LBIAAAAYAAJ}.

\bibitem[Kohlberg(1981)]{Kohlberg81-KOHTPO-6}
Lawrence Kohlberg.
\newblock \emph{The Philosophy of Moral Development: Moral Stages and the Idea of Justice}.
\newblock San Francisco : Harper \& Row, 1981.

\bibitem[Kotek et~al.()Kotek, Dockum, and Sun]{kotek_gender_2023}
Hadas Kotek, Rikker Dockum, and David Sun.
\newblock Gender bias and stereotypes in large language models.
\newblock In \emph{Proceedings of The {ACM} Collective Intelligence Conference}, pp.\  12--24. {ACM}.
\newblock ISBN 979-8-4007-0113-9.
\newblock \doi{10.1145/3582269.3615599}.
\newblock URL \url{https://dl.acm.org/doi/10.1145/3582269.3615599}.

\bibitem[Liscio et~al.(2023)Liscio, Lera-Leri, Bistaffa, Dobbe, Jonker, Lopez-Sanchez, Rodriguez-Aguilar, and Murukannaiah]{Liscio23Value}
Enrico Liscio, Roger Lera-Leri, Filippo Bistaffa, Roel~I.J. Dobbe, Catholijn~M. Jonker, Maite Lopez-Sanchez, Juan~A. Rodriguez-Aguilar, and Pradeep~K. Murukannaiah.
\newblock Value inference in sociotechnical systems.
\newblock In \emph{Proceedings of the 2023 International Conference on Autonomous Agents and Multiagent Systems}, AAMAS '23, pp.\  1774–1780, Richland, SC, 2023. International Foundation for Autonomous Agents and Multiagent Systems.
\newblock ISBN 9781450394321.

\bibitem[Liu et~al.(2025{\natexlab{a}})Liu, Cao, Wu, Qiu, Gu, Liu, and Hershcovich]{Liu25Towards}
Haijiang Liu, Yong Cao, Xun Wu, Chen Qiu, Jinguang Gu, Maofu Liu, and Daniel Hershcovich.
\newblock Towards realistic evaluation of cultural value alignment in large language models: Diversity enhancement for survey response simulation.
\newblock \emph{Information Processing \& Management}, 62\penalty0 (4):\penalty0 104099, 2025{\natexlab{a}}.
\newblock ISSN 0306-4573.
\newblock \doi{https://doi.org/10.1016/j.ipm.2025.104099}.
\newblock URL \url{https://www.sciencedirect.com/science/article/pii/S030645732500041X}.

\bibitem[Liu et~al.(2025{\natexlab{b}})Liu, Li, Gao, Cao, Xu, Wu, Hershcovich, and Gu]{liu2025demographics}
Haijiang Liu, Qiyuan Li, Chao Gao, Yong Cao, Xiangyu Xu, Xun Wu, Daniel Hershcovich, and Jinguang Gu.
\newblock Beyond demographics: Enhancing cultural value survey simulation with multi-stage personality-driven cognitive reasoning.
\newblock In \emph{Proceedings of the 2025 Conference of the The 2025 Conference on Empirical Methods in Natural Language Processing (Volume 1: Long Papers)}. Association for Computational Linguistics, 2025{\natexlab{b}}.
\newblock URL \url{https://arxiv.org/abs/2508.17855}.

\bibitem[Mercier \& Landemore(2012)Mercier and Landemore]{mercier_reasoning_2012}
Hugo Mercier and Hélène Landemore.
\newblock Reasoning is for arguing: Understanding the successes and failures of deliberation.
\newblock 33\penalty0 (2):\penalty0 243--258, 2012.
\newblock ISSN 0162-895X, 1467-9221.
\newblock \doi{10.1111/j.1467-9221.2012.00873.x}.
\newblock URL \url{https://onlinelibrary.wiley.com/doi/10.1111/j.1467-9221.2012.00873.x}.

\bibitem[Mittelstadt(2019)]{Mittelstadt19Principles}
Brent~D. Mittelstadt.
\newblock Principles alone cannot guarantee ethical {AI}.
\newblock \emph{Nat. Mach. Intell.}, 1\penalty0 (11):\penalty0 501--507, 2019.
\newblock \doi{10.1038/S42256-019-0114-4}.
\newblock URL \url{https://doi.org/10.1038/s42256-019-0114-4}.

\bibitem[Moore(2014)]{moore_democratic_2014}
Alfred Moore.
\newblock Democratic reason: Politics, collective intelligence and the rule of the many.
\newblock 13\penalty0 (2):\penalty0 e12--e15, 2014.
\newblock ISSN 1476-9336.
\newblock \doi{10.1057/cpt.2013.26}.
\newblock URL \url{https://doi.org/10.1057/cpt.2013.26}.

\bibitem[Narayanan \& Samuel(2025)Narayanan and Samuel]{Narayanan25Ethical}
Ravi Narayanan and Adebis Samuel.
\newblock Ethical challenges and bias in nlp models: A python-based investigation.
\newblock 05 2025.

\bibitem[Navigli et~al.(2023)Navigli, Conia, and Ross]{Navigli23Biases}
Roberto Navigli, Simone Conia, and Bj\"{o}rn Ross.
\newblock Biases in large language models: Origins, inventory, and discussion.
\newblock \emph{J. Data and Information Quality}, 15\penalty0 (2), June 2023.
\newblock ISSN 1936-1955.
\newblock \doi{10.1145/3597307}.
\newblock URL \url{https://doi.org/10.1145/3597307}.

\bibitem[Ouyang et~al.(2022)Ouyang, Wu, Jiang, Almeida, Wainwright, Mishkin, Zhang, Agarwal, Slama, Ray, Schulman, Hilton, Kelton, Miller, Simens, Askell, Welinder, Christiano, Leike, and Lowe]{ouyang2022traininglanguagemodelsfollow}
Long Ouyang, Jeff Wu, Xu~Jiang, Diogo Almeida, Carroll~L. Wainwright, Pamela Mishkin, Chong Zhang, Sandhini Agarwal, Katarina Slama, Alex Ray, John Schulman, Jacob Hilton, Fraser Kelton, Luke Miller, Maddie Simens, Amanda Askell, Peter Welinder, Paul Christiano, Jan Leike, and Ryan Lowe.
\newblock Training language models to follow instructions with human feedback, 2022.
\newblock URL \url{https://arxiv.org/abs/2203.02155}.

\bibitem[Park et~al.(2023)Park, O'Brien, Cai, Morris, Liang, and Bernstein]{Park23Generative}
Joon~Sung Park, Joseph O'Brien, Carrie~Jun Cai, Meredith~Ringel Morris, Percy Liang, and Michael~S. Bernstein.
\newblock Generative agents: Interactive simulacra of human behavior.
\newblock In \emph{Proceedings of the 36th Annual ACM Symposium on User Interface Software and Technology}, UIST '23, New York, NY, USA, 2023. Association for Computing Machinery.
\newblock ISBN 9798400701320.
\newblock \doi{10.1145/3586183.3606763}.
\newblock URL \url{https://doi.org/10.1145/3586183.3606763}.

\bibitem[Raza et~al.(2024)Raza, Garg, Reji, Bashir, and Ding]{Shaina24Nbias}
Shaina Raza, Muskan Garg, Deepak~John Reji, Syed~Raza Bashir, and Chen Ding.
\newblock Nbias: A natural language processing framework for bias identification in text.
\newblock \emph{Expert Systems with Applications}, 237:\penalty0 121542, 2024.
\newblock ISSN 0957-4174.
\newblock \doi{https://doi.org/10.1016/j.eswa.2023.121542}.
\newblock URL \url{https://www.sciencedirect.com/science/article/pii/S0957417423020444}.

\bibitem[Santurkar et~al.(2023)Santurkar, Durmus, Ladhak, Lee, Liang, and Hashimoto]{pmlr-v202-santurkar23a}
Shibani Santurkar, Esin Durmus, Faisal Ladhak, Cinoo Lee, Percy Liang, and Tatsunori Hashimoto.
\newblock Whose opinions do language models reflect?
\newblock In \emph{Proceedings of the 40th International Conference on Machine Learning}, volume 202 of \emph{Proceedings of Machine Learning Research}, pp.\  29971--30004. PMLR, 23--29 Jul 2023.
\newblock URL \url{https://proceedings.mlr.press/v202/santurkar23a.html}.

\bibitem[Scherrer et~al.(2023)Scherrer, Shi, Feder, and Blei]{Scherrer23Evaluating}
Nino Scherrer, Claudia Shi, Amir Feder, and David Blei.
\newblock Evaluating the moral beliefs encoded in llms.
\newblock In \emph{Advances in Neural Information Processing Systems}, volume~36, pp.\  51778--51809. Curran Associates, Inc., 2023.
\newblock URL \url{https://proceedings.neurips.cc/paper_files/paper/2023/file/a2cf225ba392627529efef14dc857e22-Paper-Conference.pdf}.

\bibitem[Sclar et~al.(2024)Sclar, Choi, Tsvetkov, and Suhr]{Sclar24Quantifying}
Melanie Sclar, Yejin Choi, Yulia Tsvetkov, and Alane Suhr.
\newblock Quantifying language models' sensitivity to spurious features in prompt design or: How {I} learned to start worrying about prompt formatting.
\newblock In \emph{The Twelfth International Conference on Learning Representations, {ICLR} 2024, Vienna, Austria, May 7-11, 2024}. OpenReview.net, 2024.
\newblock URL \url{https://openreview.net/forum?id=RIu5lyNXjT}.

\bibitem[Selbst et~al.(2019)Selbst, Boyd, Friedler, Venkatasubramanian, and Vertesi]{Selbst19Fairness}
Andrew~D. Selbst, Danah Boyd, Sorelle~A. Friedler, Suresh Venkatasubramanian, and Janet Vertesi.
\newblock Fairness and abstraction in sociotechnical systems.
\newblock In \emph{Proceedings of the Conference on Fairness, Accountability, and Transparency}, FAT* '19, pp.\  59–68, New York, NY, USA, 2019. Association for Computing Machinery.
\newblock ISBN 9781450361255.
\newblock \doi{10.1145/3287560.3287598}.
\newblock URL \url{https://doi.org/10.1145/3287560.3287598}.

\bibitem[Tao et~al.(2024)Tao, Viberg, Baker, and Kizilcec]{Tao24Cultural}
Yan Tao, Olga Viberg, Ryan~S Baker, and René~F Kizilcec.
\newblock Cultural bias and cultural alignment of large language models.
\newblock \emph{PNAS Nexus}, 3\penalty0 (9):\penalty0 pgae346, 09 2024.
\newblock ISSN 2752-6542.
\newblock \doi{10.1093/pnasnexus/pgae346}.
\newblock URL \url{https://doi.org/10.1093/pnasnexus/pgae346}.

\bibitem[Taubenfeld et~al.(2024)Taubenfeld, Dover, Reichart, and Goldstein]{taubenfeld-etal-2024-systematic}
Amir Taubenfeld, Yaniv Dover, Roi Reichart, and Ariel Goldstein.
\newblock Systematic biases in {LLM} simulations of debates.
\newblock In \emph{Proceedings of the 2024 Conference on Empirical Methods in Natural Language Processing}, pp.\  251--267, Miami, Florida, USA, November 2024. Association for Computational Linguistics.
\newblock \doi{10.18653/v1/2024.emnlp-main.16}.
\newblock URL \url{https://aclanthology.org/2024.emnlp-main.16/}.

\bibitem[Wu et~al.(2025)Wu, Sheng, Wang, Yang, Sun, Wang, Bu, and Cao]{wu2025staircaseethicsprobingllm}
Ya~Wu, Qiang Sheng, Danding Wang, Guang Yang, Yifan Sun, Zhengjia Wang, Yuyan Bu, and Juan Cao.
\newblock The staircase of ethics: Probing llm value priorities through multi-step induction to complex moral dilemmas, 2025.
\newblock URL \url{https://arxiv.org/abs/2505.18154}.

\bibitem[Yuan et~al.(2024)Yuan, Murukannaiah, and Singh]{Yuan24Right}
Jiaqing Yuan, Pradeep~K. Murukannaiah, and Munindar~P. Singh.
\newblock Right vs. right: Can llms make tough choices?
\newblock \emph{CoRR}, abs/2412.19926, 2024.
\newblock \doi{10.48550/ARXIV.2412.19926}.
\newblock URL \url{https://doi.org/10.48550/arXiv.2412.19926}.

\bibitem[Zhao et~al.(2024)Zhao, Mondal, Tandon, Dillion, Gray, and Gu]{zhao-etal-2024-worldvaluesbench}
Wenlong Zhao, Debanjan Mondal, Niket Tandon, Danica Dillion, Kurt Gray, and Yuling Gu.
\newblock {W}orld{V}alues{B}ench: A large-scale benchmark dataset for multi-cultural value awareness of language models.
\newblock In \emph{Proceedings of the 2024 Joint International Conference on Computational Linguistics, Language Resources and Evaluation (LREC-COLING 2024)}, pp.\  17696--17706, Torino, Italia, May 2024. ELRA and ICCL.
\newblock URL \url{https://aclanthology.org/2024.lrec-main.1539/}.

\end{thebibliography}
\bibliographystyle{iclr2025_conference}

\appendix
\section{Appendix}

\subsection{Preference bias revealed by DEF system}
\label{app:bias}
\begin{figure}[]
    \centering
    \subfloat[Baichuan2-13B-Chat]{
    \includegraphics[width=0.28\textwidth]{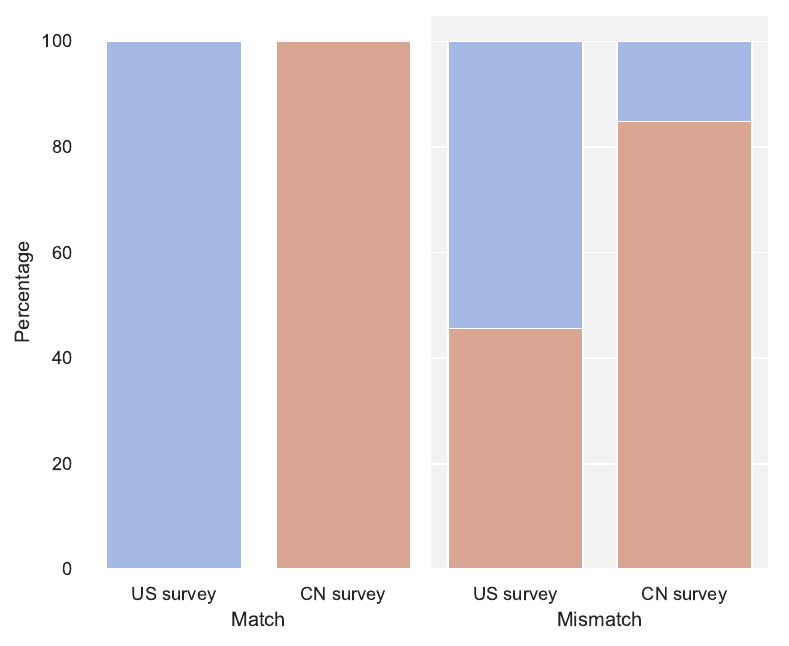}
    }
    \subfloat[ChatGLM2-6B]{
    \includegraphics[width=0.28\textwidth]{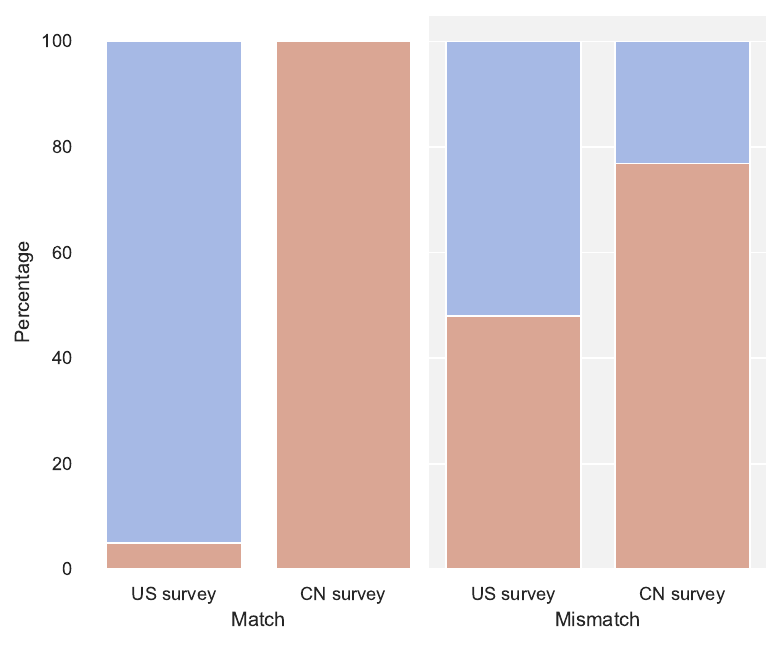}
    }  
    \subfloat[WizardLM-13B]{
    \includegraphics[width=0.34\textwidth]{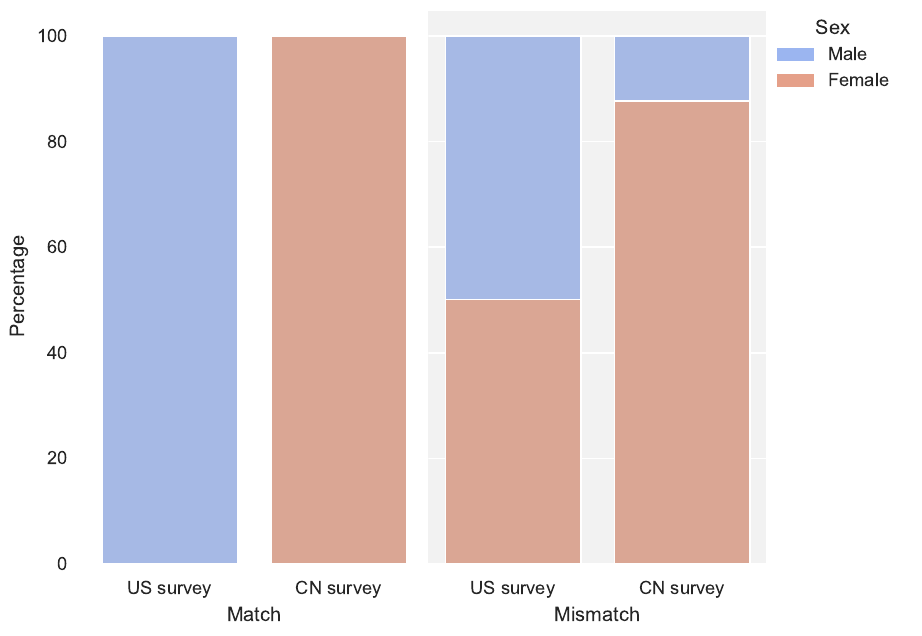}
    }
    \\
    \subfloat[Mistral-7B-Instruct]{
    \includegraphics[width=0.28\textwidth]{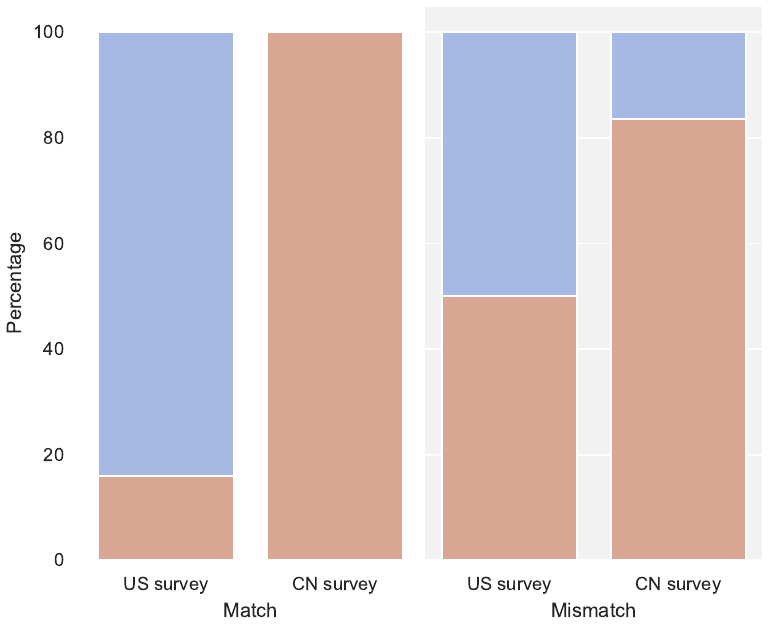}
    } 
    \subfloat[Dolphin-2.2.1-Mistral-7B]{
    \includegraphics[width=0.28\textwidth]{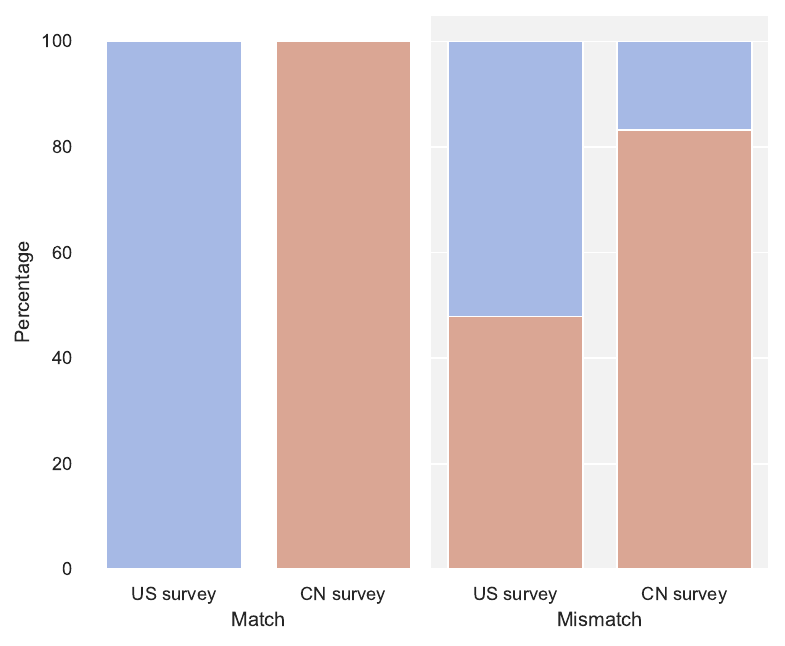}
    } 
    \subfloat[Mixtral-8x7B-Instruct]{
    \includegraphics[width=0.34\textwidth]{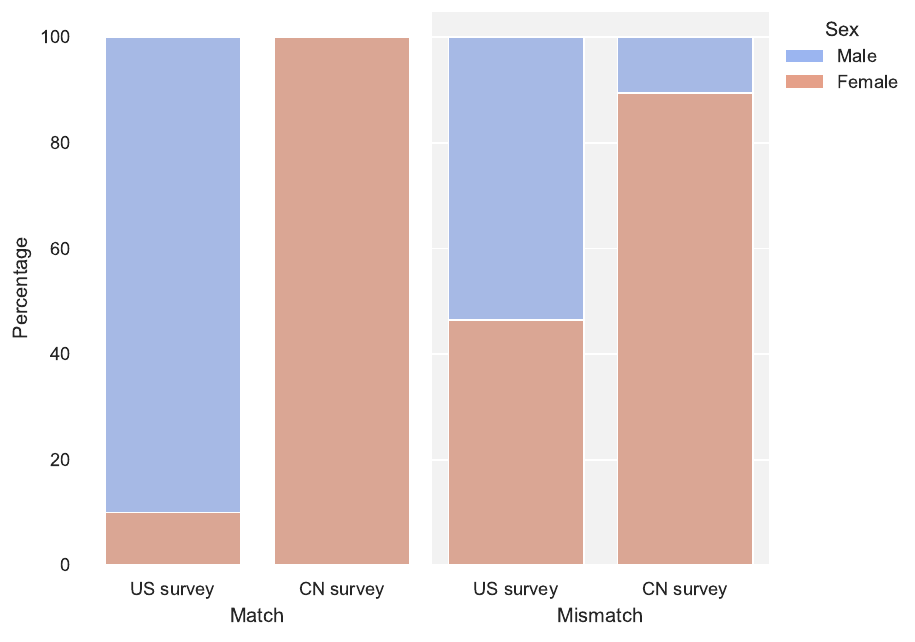}
    } 
    \\
    \subfloat[Llama-3-8B]{
    \includegraphics[width=0.28\textwidth]{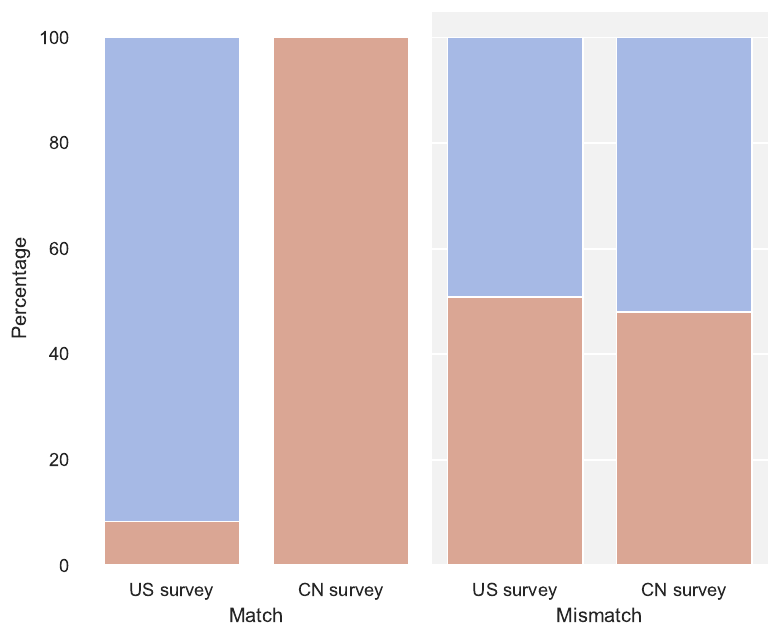}
    }
    \subfloat[Llama-3-8B-Instruct]{
    \includegraphics[width=0.28\textwidth]{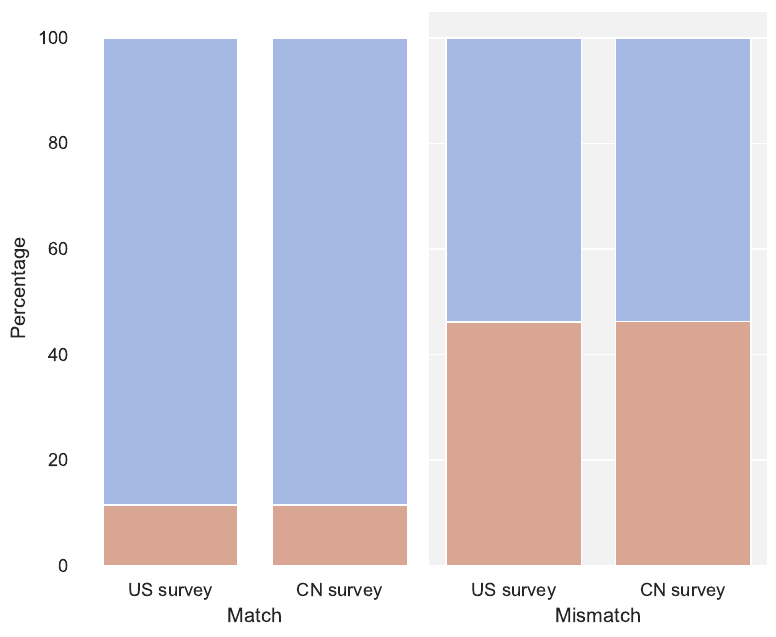}
    }
    \subfloat[Dolphin-2.9.1-Llama-3-8B]{
    \includegraphics[width=0.34\textwidth]{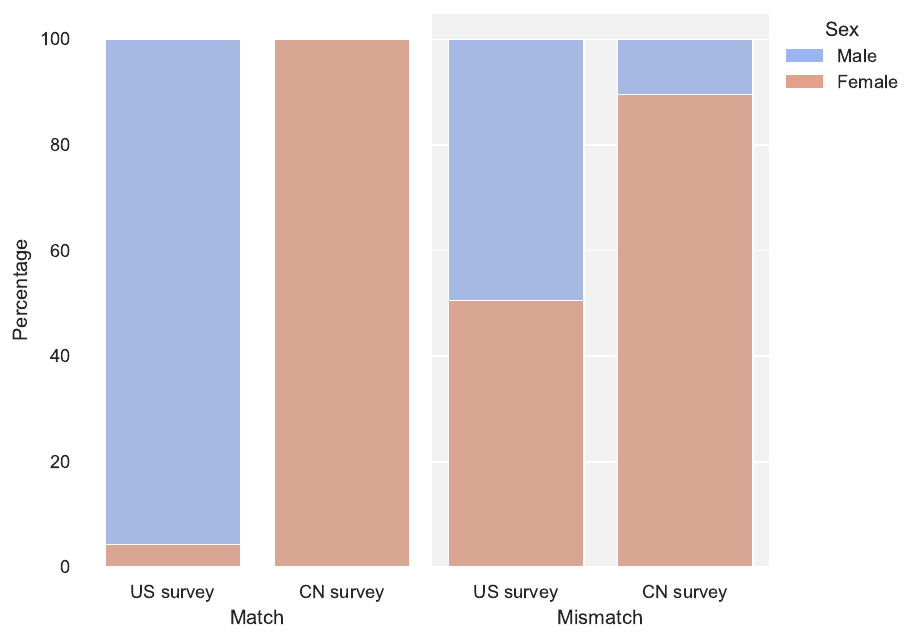}
    }
    \\
    \subfloat[Llama-3-Chinese-8B-Instruct]{
    \includegraphics[width=0.34\textwidth]{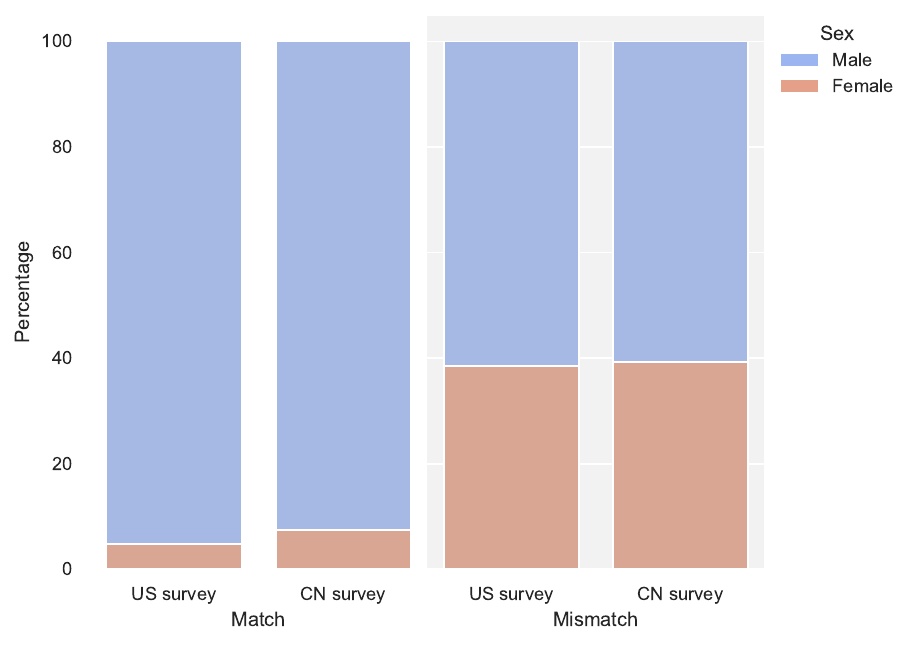}
    }
    \subfloat[Claude-3.5-Sonnet]{
    \includegraphics[width=0.34\textwidth]{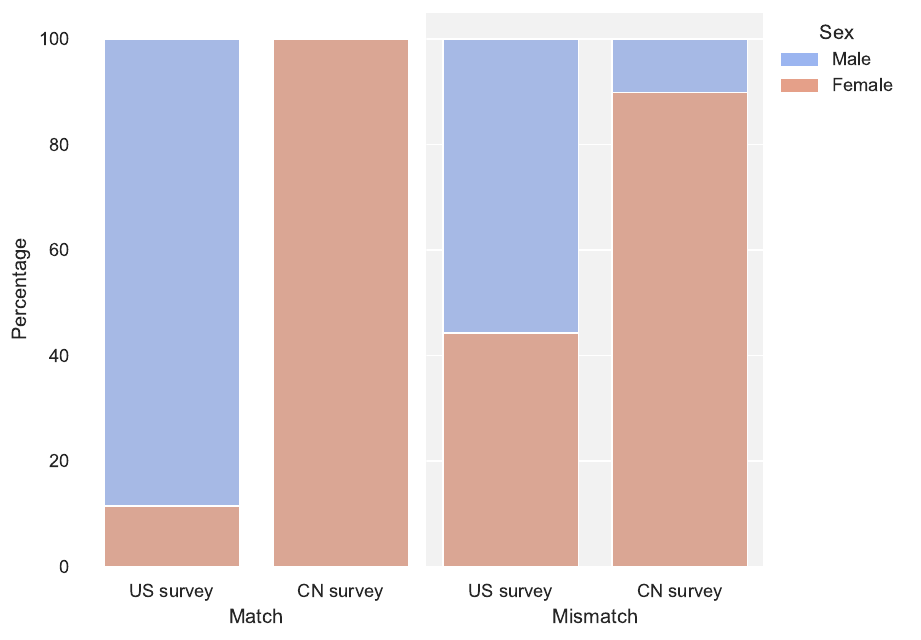}
    }
    \caption{Mismatch proportions by gender for all candidates. There is a heavy bias on male profiles in the US survey, and female profiles in the CN survey. But in general, gender bias on the mismatch profiles in the US survey is less significant than in the CN survey. \textit{Adapted from \citet{Liu25Towards}.}}
    \label{fig:gender}
    \vspace{-0.3cm}
\end{figure}

\begin{figure}[]
    \centering
    \subfloat[Baichuan2-13B-Chat]{
    \includegraphics[width=0.28\textwidth]{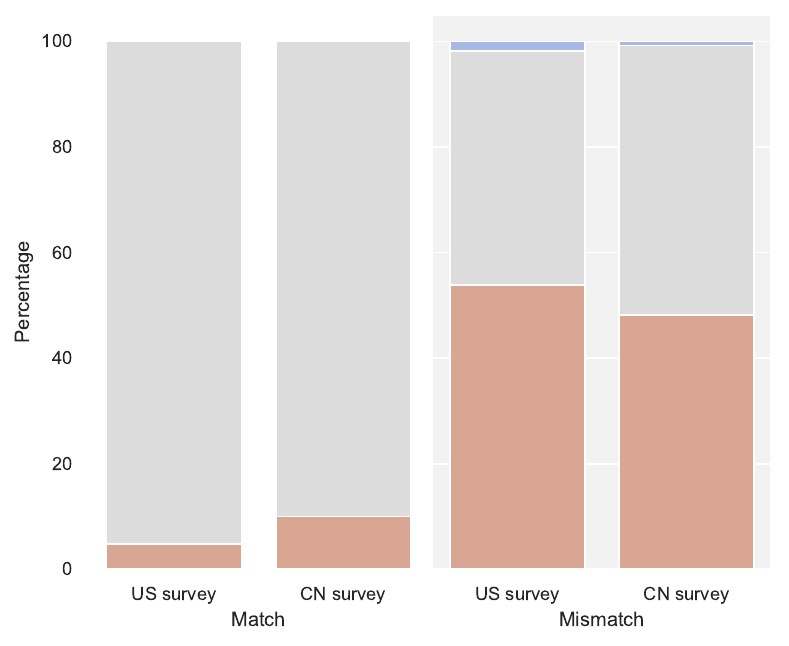}
    }
    \subfloat[ChatGLM2-6B]{
    \includegraphics[width=0.28\textwidth]{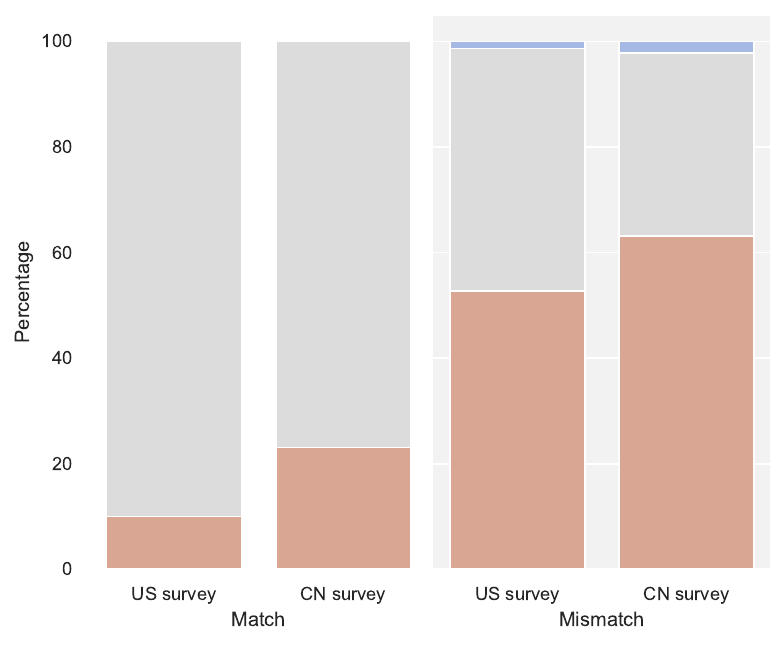}
    }  
    \subfloat[WizardLM-13B]{
    \includegraphics[width=0.34\textwidth]{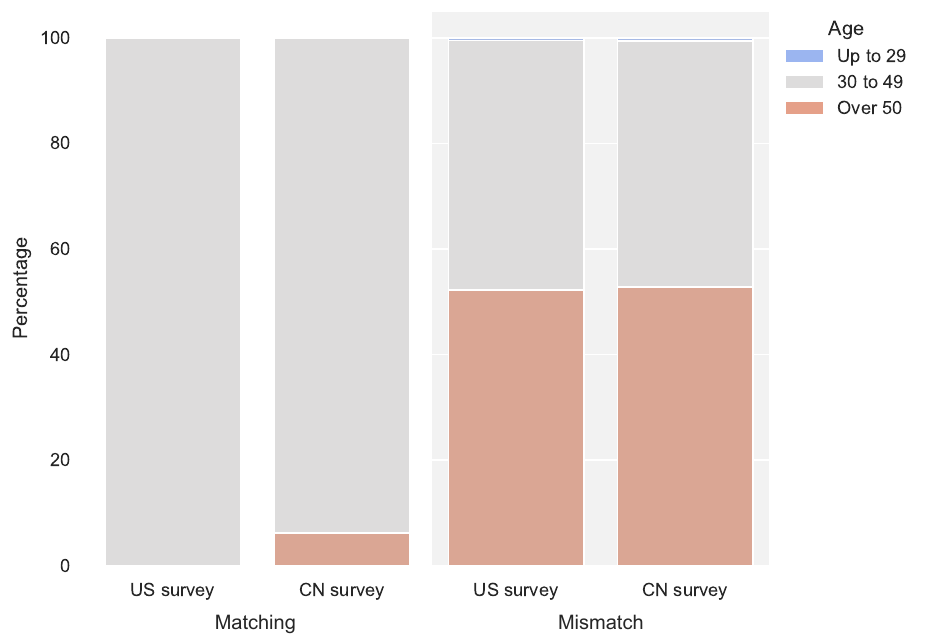}
    }
    \vspace{-0.2cm}
    \\
    \subfloat[Mistral-7B-Instruct]{
    \includegraphics[width=0.28\textwidth]{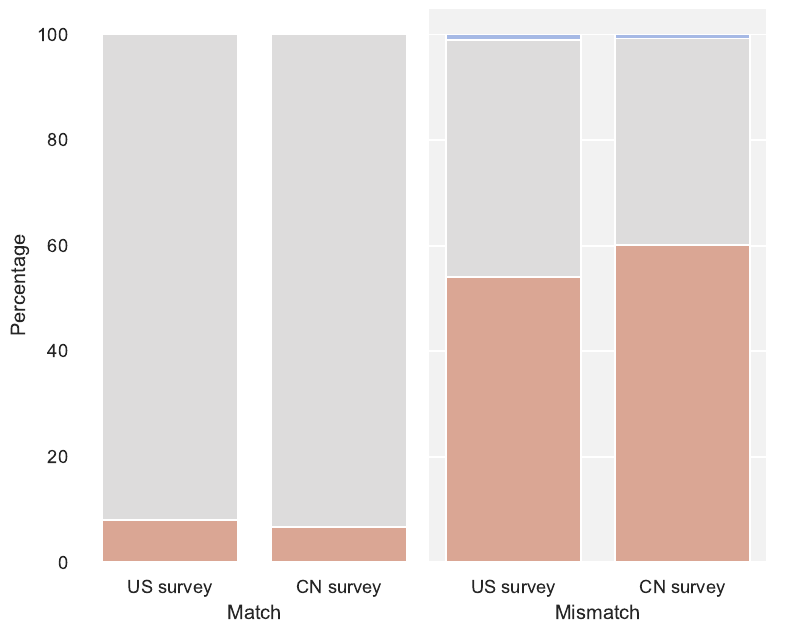}
    } 
    \subfloat[Dolphin-2.2.1-Mistral-7B]{
    \includegraphics[width=0.28\textwidth]{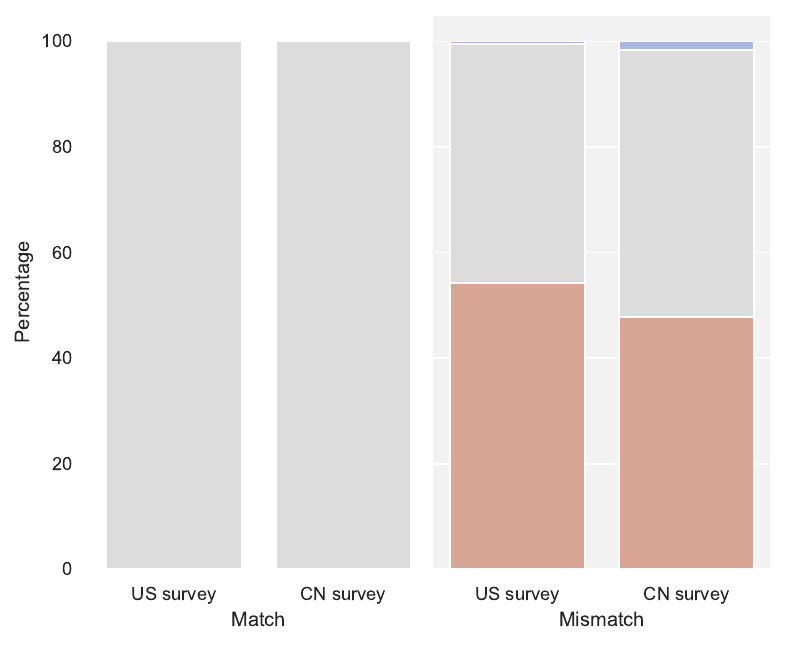}
    } 
    \subfloat[Mixtral-8x7B-Instruct]{
    \includegraphics[width=0.34\textwidth]{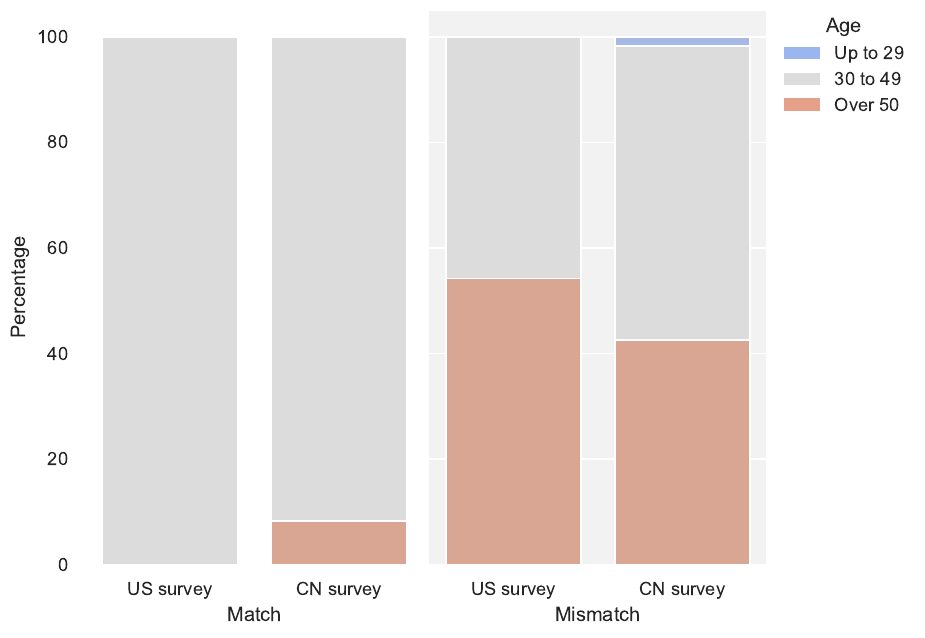}
    } 
    \vspace{-0.2cm}
    \\
    \subfloat[Llama-3-8B]{
    \includegraphics[width=0.28\textwidth]{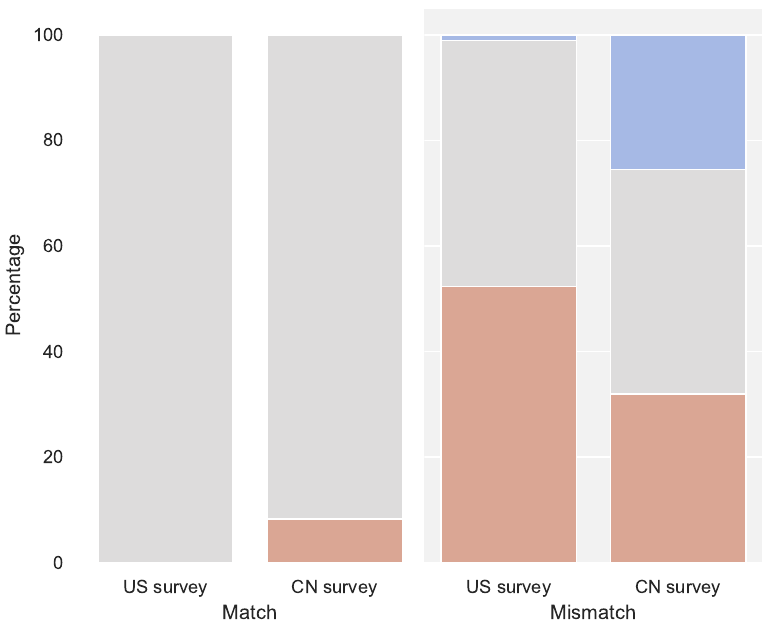}
    }
    \subfloat[Llama-3-8B-Instruct]{
    \includegraphics[width=0.28\textwidth]{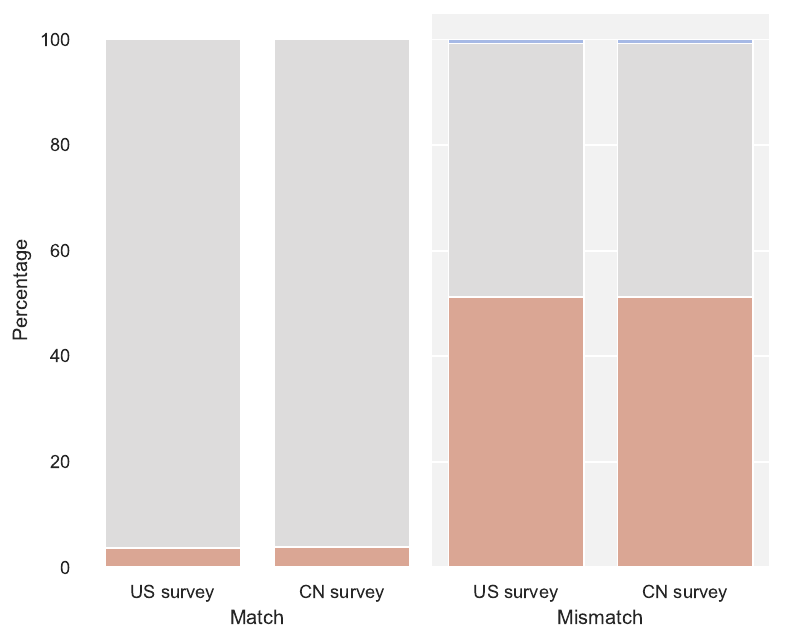}
    }
    \subfloat[Dolphin-2.9.1-Llama-3-8B]{
    \includegraphics[width=0.34\textwidth]{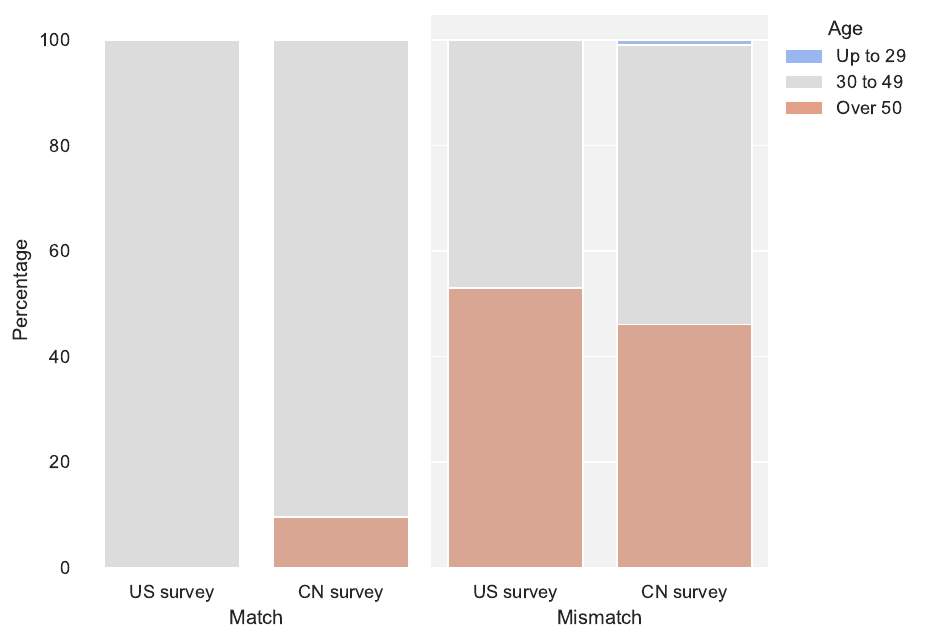}
    }
    \vspace{-0.2cm}
    \\
    \subfloat[Llama-3-Chinese-8B]{
    \includegraphics[width=0.34\textwidth]{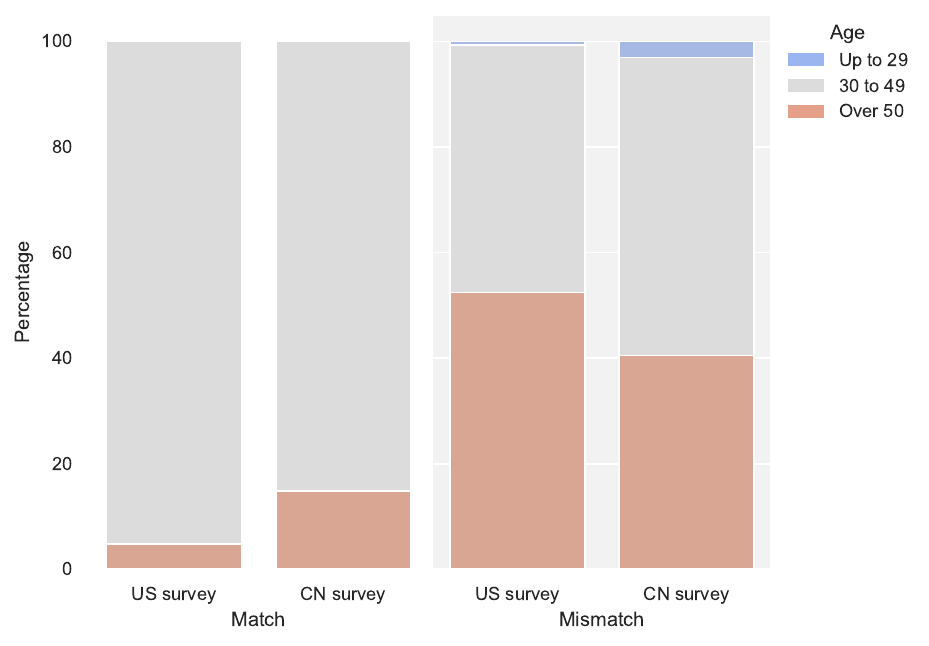}
    }
    \subfloat[Claude-3.5-Sonnet]{
    \includegraphics[width=0.34\textwidth]{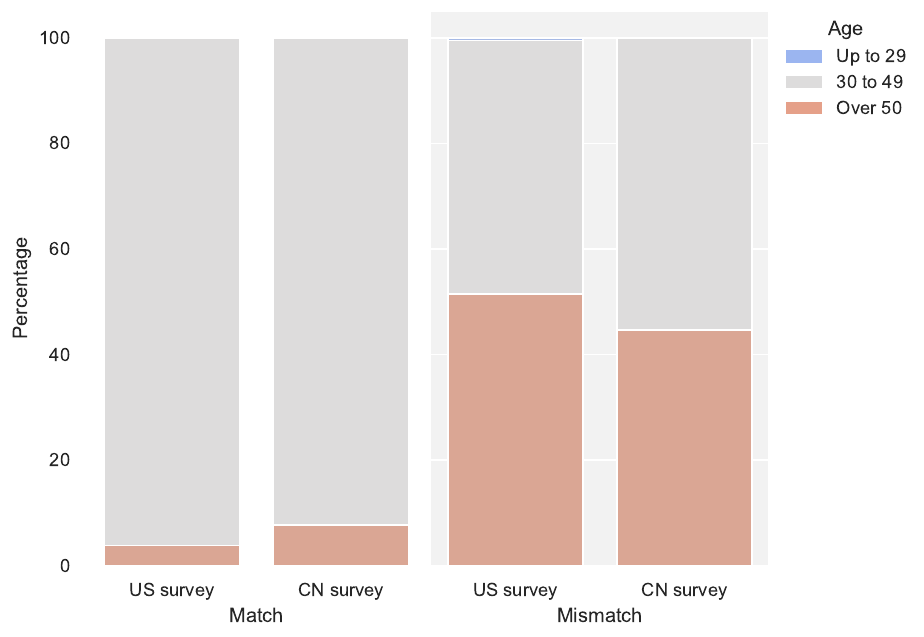}
    }
    \caption{Proportion of each age group in mismatch profiles across all candidates, following the social survey design. Both matching and mismatch profiles show a strong bias toward middle-aged characters, with under-representing preferences for characters under 29 years old. \textit{Adapted from \citet{Liu25Towards}.}}
    \label{fig:age}
    \vspace{-0.3cm}
\end{figure}

\end{document}